\documentclass[amsmath,amssymb,showkeys,superscriptaddress,floatfix,prd,11pt]{revtex4}
\usepackage{graphicx,epstopdf}
\usepackage{subfig}
\usepackage{xcolor}
\usepackage[colorlinks=true, urlcolor=blue, linkcolor=blue, citecolor=green]{hyperref}
\def\pslash{p\!\!\!\slash }
\def\uslash{u\!\!\!\slash }

\begin{document}

\title{Fate of the doubly heavy spin$-3/2$ baryons in a dense medium}
\author{N. Er}%
\email[]{nuray@ibu.edu.tr}
\affiliation{Department of Physics, Abant \.{I}zzet Baysal University,
G\"olk\"oy Kamp\"us\"u, 14980 Bolu, Turkey}
\author{K. Azizi}%
\email[]{kazizi@dogus.edu.tr}
\affiliation{Department of Physics, Dogus University, Acibadem-Kadikoy, 34722 
Istanbul, Turkey}
\affiliation{School of Particles and Accelerators, Institute for Research in Fundamental Sciences (IPM) P.O. Box 19395-5531, Tehran, Iran}

\date{\today}
 
\begin{abstract}
We investigate the behavior  of the doubly heavy spin$-3/2$ baryons in cold nuclear matter. In particular, we study the variations of the spectroscopic parameters of the  ground state $\Xi^*_{QQ'}$ and $\Omega^*_{QQ'}$ particles,  with $ Q $ and $ Q' $ being $b  $ or $ c $ quark, with respect to the changes in the density of the nuclear medium.  We find the shifts on the parameters under question at saturation medium density compared to their vacuum values. It is observed that the parameters of the $\Xi^*_{QQ'}$ states containing two heavy quarks and one up or down quark are affected by the medium, considerably. The parameters of the  $\Omega^*_{QQ'}$ states   containing two heavy quarks and one strange quark, however,  do not show any sensitivity to the density of the cold nuclear medium. We also discuss the variations of the vector self-energy at each channel with respect to the changes in the density. The negative shifts in the mass of $\Xi^*_{QQ'}$ states due to nucleons in the medium can be used to study the doubly heavy baryons' interactions with the nucleons. The results obtained can also be  used in analyses of the  results of the  future in-medium experiments.
\end{abstract}
\keywords{Heavy Baryons, Nuclear Matter, In-medium Sum Rules}

\maketitle

\section{Introduction}
The investigation of the hadronic properties under extreme conditions, without doubt, is one of the main goals of the quantum chromodynamics (QCD) and  hadron physics. Such investigations will help us gain valuable information on the internal structures of hadrons, their probable melting at a critical temperature/density, probable transition to the quark-gluon-plasma (QGP) as a possible new phase of matter, structure of the dense astrophysical objects like neutron stars, analyses of the results of the  heavy-ion collision and in-medium experiments as well as the perturbative and non-perturbative natures of QCD as the theory of one of the fundamental interactions of nature.  A class of hadrons that deserves investigation in nuclear medium is doubly heavy baryons. Such explorations can help us study the interactions of the doubly heavy baryons with nucleons. 

The existence of $\Xi^{(*)}_{QQ'}$ and $\Omega^{(*)}_{QQ'}$ with different spin-parity and heavy-light quarks contents is a natural outcome of the quark model. Though, their properties have been widely investigated in theory and their nature and internal structure have been stabilized theoretically, our experimental knowledge on these states  are limited to the state   $\Xi_{cc}$ with spin$-1/2$. The  existence of   $\Xi^+_{cc}$ was firstly reported by SELEX Collaboration in 2002 \cite{PhysRevLett.89.112001} and confirmed by the same collaboration \cite{OCHERASHVILI200518} in 2005,  however,  it was not been later confirmed by other experimental groups.  The SELEX Collaboration reported the mass $ 3518.7\pm 1.7 ~\text{MeV}/c^2 $ for this state. The theoretical studies, however, mainly predicted its mass to lie in the interval $ [3500-3720] ~\text{MeV}/c^2 $  \cite{ALIEV201259,Gershtein2000,PhysRevD.66.014502,PhysRevD.91.094030,PhysRevD.90.094507,PhysRevD.95.116012,PhysRevD.92.034504,PhysRevD.96.034511,PhysRevD.93.094002,PhysRevD.92.076008,PhysRevD.90.094007,Wang:2010hs,PhysRevD.78.094007,Roberts:2007ni,PhysRevD.66.014008,Kiselev:2001fw}. In \cite{ALIEV201259}, for instance, its mass was predicted to be  $ 3.72\pm 0.20 ~\text{GeV}/c^2 $. Recently, the LHCb Collaboration has  reported the observation of doubly heavy baryon $ \Xi^{++}_{cc} $ via the decay mode $ \Lambda_c^+ K^- \pi^+\pi^+$ with mass $ 3621.40\pm 0.72 (\text{stat.})\pm 0.27 (\text{syst.})\pm 0.14 (\Lambda_c^+)~\text{MeV}/c^2 $ \cite{PhysRevLett.119.112001}. Although it is still under debt why the observed mass by LHCb differs considerably from the  SELEX result, the existence of a $\Xi_{cc}$  doubly charmed baryon is now on a more solid ground, experimentally. The observation of LHCb has increased the attentions to doubly heavy baryons.

The spectroscopic parameters  and production of the doubly heavy baryons have been widely investigated via different methods and approaches in vacuum \cite{ALIEV201259,Gershtein2000,PhysRevD.66.014502,PhysRevD.91.094030,PhysRevD.90.094507,PhysRevD.95.116012,PhysRevD.92.034504,PhysRevD.96.034511,PhysRevD.93.094002,PhysRevD.92.076008,PhysRevD.90.094007,Wang:2010hs,PhysRevD.78.094007,Roberts:2007ni,PhysRevD.66.014008,Kiselev:2001fw,ALIEV2012149,0954-3899-40-6-065003,ZHAO2017349,MARTYNENKO2008317,Wang:2010vn,ALBUQUERQUE2010217,refId0,Chen2014,MA2015463,Shah:2016vmd,PhysRevD.94.074003,Kiselev2018,PhysRevD.96.114006,PhysRevD.97.054008,PhysRevD.98.113004,Yu:2018com,PhysRevD.98.094021}.  The studies devoted to their investigation in medium, however,  are very limited \cite{PhysRevD.97.074003,spi1half,Wang2012}. In this study, we explore the properties of the  doubly heavy  $\Xi^*_{QQ'}$ and $\Omega^*_{QQ'}$  spin$-3/2$ baryons in cold nuclear matter. We use the in-medium QCD sum rules to study the variations of the mass and residue of these baryons with respect to the density when $\rho_N$  varies in the interval  $[0-1.5] \times\rho_N^{sat}  $.    We also discuss the variations of vector self energies of the states under considerations with respect to the changes in nuclear matter density. We comprehensively compare our results on the spectroscopic parameters of the considered baryons with the existing vacuum predictions by switching  $\rho_N\rightarrow 0$. Our results may help  groups aiming to produce and study the properties of these hadrons in dense medium.  The in-medium properties of light baryons have been investigated in Refs. \cite{Azizi2014,PhysRevD.92.054026,PhysRevC.94.065201,Drukarev:2013kga,PhysRevC.88.035208}. The fate of the heavy baryons with single heavy quark in dense medium was studied in \cite{PhysRevC.85.045204,AZIZI2017147,AZIZI2018422,PhysRevD.99.014026}. We investigated the properties of  the exotic $ X(3872) $ state, containing the charm and 
anticharm quarks and assuming a  diquark-antidiquark  organization for its internal structure, propagating in a dense medium \cite{AZIZI2018151}. We found that the mass of the $ X(3872) $ state decreases with increasing the density of the medium, considerably. We shall also refer to the pioneering works \cite{PhysRevLett.66.2851,PhysRevC.46.R34}, where the in-medium QCD sum rules with tensor condensates were presented  for the first time.

In next section, we use the in-medium QCD sum rules to obtain the expressions of the in-medium mass, residue and vector self energy as  functions of the density as well as the Borel mass parameter and in-medium continuum threshold, appearing after some transformations with the aim of killing the effects of higher  resonances and continuum states. We impose the medium effects via modifying the quark, gluon and mixed condensates in terms of density as well as considering extra operators appearing at finite density. In section III, we numerically analyze the obtained in-medium sum rules for the physical quantities under consideration and discuss their variations with respect to the density. By giving the percentage of the pole contribution, we turn off the density and obtain the results in vacuum. We compare the results obtained in this limit with existing results of the previous studies via various methods. Section IV, is devoted to the discussions and comments. We reserve the appendix  to represent the expressions of the spectral densities obtained.

\section{Formalism}
We use the in-medium QCD sum rule formalism to find the expressions of the mass and residue of the doubly heavy spin$-3/2$ baryons in terms of the cold nuclear matter density as well as the parameters of the model. But, before proceeding, let us have a comment on the classifications of the ground state doubly heavy baryons according to the quark model.  In the case of the ground state  doubly  heavy
baryons with two identical heavy quarks, i.e., the baryons $\Xi^{(*)}_{QQ}$ and
$\Omega^{(*)}_{QQ}$, the pair of heavy quarks form a diquark  with total
spin of 1. Here, baryons with star refer to the baryons of spin--3/2 and those without star to the spin--1/2 baryons.  By addition of  the spin--1/2 of the  light quark, we get two states with
total spin 1/2 and 3/2.  For these states, the interpolating currents
should be symmetric with respect to the exchange of the  heavy
quark fields. In the case of  the states containing two different heavy quarks, in addition
to the previous possibility, i.e., a diquark with total spin  1, 
the diquark can also have the total spin zero, leading to the total
spin 1/2 for these states. The interpolating currents of these states, which are denoted by $\Xi'_{bc}$  and $\Omega'_{bc}$,  are
anti-symmetric with respect to the exchange of the  heavy quark fields. In the
present work, as we previously mentioned, we deal only with spin--3/2 states. The starting point is to consider an appreciate in-medium two-point correlation function as the building block of the formalism:
\begin{equation}\label{corre}
\Pi_{\mu\nu}(p)=i\int{d^4 xe^{ip\cdot x}\langle\psi_0|\mathcal{T}[J_{\mu}(x)\bar{J}_{\nu}(0)]|\psi_0\rangle},
\end{equation}
where  $p$ is the external four-momentum of the double heavy baryon, $|\psi_0\rangle$ is the parity and time-reversal symmetric ground state of the nuclear medium, $\mathcal{T}$ is the time ordering operator and $J_{\mu}(x)$ is the interpolating current of the doubly heavy spin$-3/2$ baryons. The nuclear medium is parametrized by the medium density and the matter four-velocity $u_{\mu}$. The colorless interpolating field, $J_{\mu}(x)$, of $\Xi^{*}_{QQ'}$ and $\Omega^{*}_{QQ'}$ particles in terms of heavy and light quarks can be written in a compact form as,
\begin{equation}
\label{current}
J_{\mu}(x)=\frac{1}{\sqrt{3}}\epsilon_{abc}\Big\{[q^{aT}C\gamma_{\mu}Q^b]Q'^{c} + [q^{aT}C\gamma_{\mu}Q'^{b}]Q^{c} + [Q^{aT}C\gamma_{\mu}Q'^{b}]q^{c} \Big\},
\end{equation}
where $\epsilon_{abc}$ is the anti-symmetric Levi-Civita tensor with $a, b, c$ being color indices, $q$ is the light quark flavor, and  $Q$ and $Q'$ are the heavy quark flavors. In the above current, $T$ represents a transpose in Dirac space and $C$ is the charge-conjugation operator. In table I, the quark flavors of the doubly heavy spin$-3/2$ baryons are presented.
\begin{table}[htp]
	\addtolength{\tabcolsep}{10pt}
	\begin{center}
\begin{tabular}{c|c|c|ccc}
	   \hline\hline
	   & $q$ &  $Q$	& $Q'$\\
	   \hline\hline
	  $\Xi^*_{QQ'}$ & $u$ / $d$ & $b$ / $c$ & $b$ / $c$ \\
	  $\Omega^*_{QQ'}$ & $s$  & $b$ / $c$ & $b$ / $c$  \\
	   	   \hline\hline
\end{tabular}
\end{center}
\caption{The quark flavors of the doubly heavy spin$-3/2$ baryons.}
	\label{table}
\end{table}

According to the standard prescriptions of the method, the correlation function in Eq.~(\ref{corre}) is calculated in two different ways: In the physical or phenomenological side, the calculations are carried out in terms of the hadronic parameters like mass and residue. In the theoretical or QCD side, the calculations are performed in terms of quarks and gluons and their mutual interactions and their interactions with nuclear matter. The later is parametrized in terms of the in-medium condensates of different dimensions. By matching the coefficients of the same structures from two sides, through a dispersion relation, the sum rules for the physical quantities in momentum states are obtained. To enhance the ground states and suppress the contributions of the higher states and continuum, a Borel transformation is applied and a continuum subtraction is performed.

To obtain the physical side, a complete set of doubly heavy baryons' state with the same quantum numbers as the interpolating current of the same state is inserted into the correlation function in Eq.~(\ref{corre}). Performing the integral over four-x, we consequently get
\begin{equation}
\label{phys}
\Pi^{Phys}_{\mu\nu} (p) = -\frac{\langle\psi_0|J_{\mu}(0)|B_{DH}(p^*,s)\rangle \langle B_{DH}(p^*,s)|\bar{J}_{\nu}(0)|\psi_0\rangle}{p^{*2}-m^{*2}_{DH}} + ...,
\end{equation}
where $p^{*}$ and $m^{*}_{DH}$ are in-medium four momentum and  the modified mass of the $|B_{DH}(p^*,s)\rangle $ doubly heavy (DH) baryonic state with spin $s$ in cold nuclear matter, respectively. In the above relation, the in-medium mass is $m^{*}_{DH}=m_{DH}+\Sigma_S$, with  $m_{DH}$ being the vacuum mass  and  $\Sigma_S$ being the scalar self-energy or the mass shift of the baryon due to nuclear medium. The dots in Eq.~(\ref{phys}) represent the contributions arising from the higher resonances and continuum states. The matrix elements in the  numerator of Eq.~(\ref{phys}) are parametrized in terms of the in-medium residue or coupling strength of the doubly heavy baryon, $\lambda^{*}_{DH}$, and the Rarita-Schwinger spinor $u_{\mu}(p^*,s)$
\begin{eqnarray}
\label{Rarita}
\langle\psi_0|J_{\mu}(0)|B_{DH}(p^*,s)\rangle & = &   \lambda^{*}_{DH} u_{\mu}(p^*,s), \nonumber \\
\langle B_{DH}(p^*,s)|\bar{J}_{\nu}(0)|\psi_0\rangle & = & \bar{\lambda}^{*}_{DH} \bar{u}_{\nu}(p^*,s).
\end{eqnarray}
After inserting Eq.~(\ref{Rarita}) into Eq.~(\ref{phys}) and summing over the spins of the $B_{DH}$ heavy baryonic states, the physical side of the correlation function is developed. The summation over the Rarita-Schwinger spinors is performed as
\begin{align}\label{RaritaRarita}
\sum_{s}u_{\mu}(p^*,s)\bar u_{\nu}(p^*,s)=-\Big(\pslash^*+m^{*}_{DH}\Big)\Big[g_{\mu\nu} - \frac{1}{3}\gamma_{\mu}\gamma_{\nu}-\frac{2\,p^*_{\mu}p^*_{\nu}}{3m^{*2}_{DH}} + \frac{p^*_{\mu}\gamma_{\nu}-p^*_{\nu}\gamma_{\mu}}{3m^{*}_{DH}}\Big],
\end{align}
and we have the correlation function  in the following form
\begin{align}
\label{}
\Pi^{Phys}_{\mu\nu}(p)=\frac{\lambda^{*}_{DH}\bar{\lambda}^{*}_{DH}(\pslash^*+m^{*}_{DH})}{p^{*2}-m^{*2}_{DH}} \Big[g_{\mu\nu} - \frac{1}{3}\gamma_{\mu}\gamma_{\nu}-\frac{2\,p^*_{\mu}p^*_{\nu}}{3m^{*2}_{DH}} + \frac{p^*_{\mu}\gamma_{\nu}-p^*_{\nu}\gamma_{\mu}}{3m^{*}_{DH}}\Big]+ ...,
\end{align}
where $p^*_{\mu}=p_{\mu}-\Sigma_{\mu,\upsilon}$ with $\Sigma_{\mu,\upsilon}$ being the vector self-energy of the baryon. The vector self energy dependens on the four momentum of the particle and the four velocity of the medium in the form: $\Sigma_{\mu,\upsilon}=\Sigma_{\upsilon}u_{\mu}+\Sigma'_{\upsilon}p_{\mu}$ with the constants $\Sigma_{\upsilon}$ and $\Sigma'_{\upsilon}$. In the mean field approximation, $\Sigma_S$ and $\Sigma_{\upsilon}$ are real and momentum independent and $\Sigma'_{\upsilon}$ is taken to be identically zero, for details see \cite{Cohen:1994wm}.

In this study, the QCD sum rule approach in vacuum is extended into the finite density problem of doubly heavy baryons propagating in a cold nuclear medium. Unlike the vacuum, there are two independent Lorentz vectors in medium: four-momentum of the particle $p_{\mu}$ and the four-velocity of the nuclear matter $u_{\mu}$. We work at the rest frame of the medium $u_{\mu}(1, 0)$. Considering the above explanations, the physical side of the correlation function is written in terms of the new and extra structures as,
\begin{eqnarray}\label{piPhys}
\label{}
\Pi^{Phys}_{\mu\nu}(p)&=&\frac{\lambda^{*}_{DH}\bar{\lambda}^{*}_{DH}}{p^{2}+\Sigma_{\upsilon}^2-2p_0\Sigma_{\upsilon}-m^{*2}_{DH}} \big(\pslash-\Sigma_{\upsilon}\uslash+m^{*}_{DH}\big) \Big\{g_{\mu\nu} - \frac{1}{3}\gamma_{\mu}\gamma_{\nu} -\frac{2}{3\, m^{*2}_{DH}} \Big[p_{\mu}p_{\nu} \nonumber \\
&-&\Sigma_{\upsilon}p_{\mu}u_{\nu}-\Sigma_{\upsilon}u_{\mu}p_{\nu}+\Sigma^2_{\upsilon}u_{\mu}u_{\nu}\Big] +\frac{1}{3\, m^{*}_{DH}} \Big[p_{\mu}\gamma_{\nu}-\Sigma_{\upsilon}u_{\mu}\gamma_{\nu}-p_{\nu}\gamma_{\mu}+\Sigma_{\upsilon}u_{\nu}\gamma_{\mu}\Big]\Big\}+ ..., \nonumber \\
\end{eqnarray}
where the variable $p_0=p\cdot u$ is used for the energy of the quasi-particle state.

At this stage we have two inevitable problems: i) The Lorentz structures seen in Eq.~(\ref{piPhys}) are not all independent, ii) the interpolating current of the doubly heavy baryons, $J_{\mu}(0)$, couples to both the spin$-1/2$ and spin$-3/2$ states. The contributions coming from  spin$-1/2$ states are unwanted in our case and we need to separate only the spin$-3/2$ states contributions in further calculations. To exclude the pollution of spin$-1/2$ states and get independent structures, we treat as follows. The matrix element of $J_{\mu}(0)$ sandwiched between the spin$-1/2$ states and the ground state of the cold nuclear matter is parametrized as
\begin{equation}
\label{spin12}
\langle\psi_0|J_{\mu}(0)|\frac{1}{2}(p)\rangle=\Big[\zeta_1 p_{\mu}+\zeta_2 \gamma_{\mu}\Big] u(p),
\end{equation}
where $\zeta_1$ and $\zeta_2$ are some constants. Imposing the condition $J_{\mu} \gamma^{\mu}=0$, we immediately obtain  $\zeta_1$ in terms of $\zeta_2$. Hence, we have
\begin{equation}
\label{spin12spin12}
\langle\psi_0|J_{\mu}(0)|\frac{1}{2}(p)\rangle=\zeta_2 \Big[\gamma_{\mu} - \frac{4}{m^{*}_{\frac{1}{2}}} p_{\mu}\Big] u(p),
\end{equation}
where $  m^{*}_{\frac{1}{2}}$ stands for the in-medium mass of the doubly heavy spin$-1/2$ particles. As is seen from Eq.~(\ref{spin12spin12}), the unwanted spin$-1/2$ contributions are proportional to $\gamma_{\mu}$ and $p_{\mu}$. In the case of current with the index $ \nu $ in Eq. (\ref{corre}), these contributions are proportional to $\gamma_{\nu}$ and $p_{\nu}$.
 In the present work, in order to get independent structures, the Dirac matrices are ordered in the form $\gamma_{\mu}\pslash\uslash\gamma_{\nu}$. To eliminate the pollution of spin$-1/2$ states, the terms proportional to  $p_{\mu}$ and $p_{\nu}$ as well as those beginning with $\gamma_{\mu}$  and ending with $\gamma_{\nu}$ are set to zero. After these procedures, we obtain the physical side of the correlation function in terms of the structures, which are independent and give contributions only to spin$-3/2$ states, as
\begin{eqnarray}
\label{}
\Pi^{Phys}_{\mu\nu}(p)&=&\frac{\lambda^{*}_{DH}\bar{\lambda}^{*}_{DH}}{p^{2}-\mu^2_{DH}} \Big\{m^{*}_{DH} g_{\mu\nu} + g_{\mu\nu}\pslash-\Sigma_{\upsilon}g_{\mu\nu}\uslash + \frac{4\,\Sigma^2_{\upsilon}}{3\,m^{*}_{DH}}u_{\mu}u_{\nu} \nonumber \\
&+&  \frac{2\,\Sigma^2_{\upsilon}}{3\,m^{*2}_{DH}}u_{\mu}u_{\nu}\pslash - \frac{2\,\Sigma^3_{\upsilon}}{3\,m^{*2}_{DH}}u_{\mu}u_{\nu}\uslash\Big\}+ ...,
\end{eqnarray}
where $\mu^2_{DH}=m^{*2}_{DH}-\Sigma_{\upsilon}^2+2p_0\Sigma_{\upsilon}$. In this stage, we apply the Borel transformation on the variable $p^2$ with the aiming of suppressing the contributions of the higher states and continuum. As a result, the physical side of the correlation function is
\begin{eqnarray}
\label{}
\Pi^{Phys}_{\mu\nu}(p)=&&\lambda^{*}_{DH}\bar{\lambda}^{*}_{DH} e^{-\mu^2_{DH}/M^2} \Big\{m^{*}_{DH} g_{\mu\nu} + g_{\mu\nu}\pslash-\Sigma_{\upsilon}g_{\mu\nu}\uslash + \frac{4\,\Sigma^2_{\upsilon}}{3\,m^{*}_{DH}}u_{\mu}u_{\nu} \nonumber \\
&+&  \frac{2\,\Sigma^2_{\upsilon}}{3\,m^{*2}_{DH}}u_{\mu}u_{\nu}\pslash - \frac{2\,\Sigma^3_{\upsilon}}{3\,m^{*2}_{DH}}u_{\mu}u_{\nu}\uslash\Big\}+ ...,
\end{eqnarray}
where $M^2$ is the Borel parameter to be fixed in the next section.

The next step is to calculate the QCD side of the correlation function in Eq.~(\ref{corre}) using the interpolating current in Eq.~(\ref{current}). Thus, in Eq.~(\ref{corre}) contracting the heavy and light quarks fields, for the case $Q\neq Q'$, we find the QCD side of the in-medium correlation function in terms of the quark propagators as
\begin{eqnarray}\label{correPi1}
\Pi^{QCD}_{\mu\nu} (p) &=&\frac{1}{3} \epsilon_{abc}\epsilon_{a'b'c'} \int d^4 x e^{ip\cdot x} \langle\psi_0|\Big\{- S^{cb'}_Q \gamma_{\nu} \tilde{S}^{aa'}_{Q'} \gamma_{\mu} S^{bc'}_q  - S^{ca'}_{Q} \gamma_{\nu} \tilde{S}^{bb'}_{q} \gamma_{\mu} S^{ac'}_{Q'} - S^{ca'}_{Q'}  \gamma_{\nu} \tilde{S}^{bb'}_{Q} \gamma_{\mu} S^{ac'}_q \nonumber \\
&-& S^{ca'}_{Q'}  \gamma_{\nu} \tilde{S}^{bb'}_{Q} \gamma_{\mu} S^{ac'}_q -  S^{ca'}_{q}  \gamma_{\nu} \tilde{S}^{bb'}_{Q'} \gamma_{\mu} S^{ac'}_Q  -  S^{cb'}_{q}  \gamma_{\nu} \tilde{S}^{aa'}_{Q} \gamma_{\mu} S^{bc'}_{Q'}  - S^{cc'}_{Q'} Tr \Big[S^{ba'}_{Q}  \gamma_{\nu} \tilde{S}^{ab'}_{q} \gamma_{\mu} \Big] \nonumber \\
&-&  S^{cc'}_{q} Tr \Big[S^{ba'}_{Q'}  \gamma_{\nu} \tilde{S}^{ab'}_{Q} \gamma_{\mu} \Big] - S^{cc'}_{Q} Tr \Big[S^{ba'}_{q}  \gamma_{\nu} \tilde{S}^{ab'}_{Q'} \gamma_{\mu} \Big]\Big\}|\psi_0\rangle,
\end{eqnarray}
where $S^{ij}_{Q(q)}$ is the heavy(light) quark propagator in the coordinate space and $\tilde{S}^{ij}_{Q(q)} = CS^{ijT}_{Q(q)} C$.  While for the case $Q=Q'$, considering the extra contractions coming form the identical particles, we get  
\begin{eqnarray}\label{correPi2}
\Pi^{QCD}_{\mu\nu} (p) &=&\frac{1}{3} \epsilon_{abc}\epsilon_{a'b'c'} \int d^4 x e^{ip\cdot x} \langle\psi_0|\Big\{S^{ca'}_Q \gamma_{\nu} \tilde{S}^{ab'}_{Q} \gamma_{\mu} S^{bc'}_q - S^{ca'}_Q \gamma_{\nu} \tilde{S}^{bb'}_{Q} \gamma_{\mu} S^{ac'}_q + S^{ca'}_Q \gamma_{\nu} \tilde{S}^{ab'}_{q} \gamma_{\mu} S^{bc'}_Q \nonumber \\
&-& S^{ca'}_Q \gamma_{\nu} \tilde{S}^{bb'}_{q} \gamma_{\mu} S^{ac'}_Q - S^{cb'}_Q \gamma_{\nu} \tilde{S}^{aa'}_{Q} \gamma_{\mu} S^{bc'}_q + S^{cb'}_Q \gamma_{\nu} \tilde{S}^{ba'}_{Q} \gamma_{\mu} S^{ac'}_q - S^{cb'}_Q \gamma_{\nu} \tilde{S}^{aa'}_{q} \gamma_{\mu} S^{bc'}_Q \nonumber \\
&+& S^{cb'}_Q \gamma_{\nu} \tilde{S}^{ab'}_{q} \gamma_{\mu} S^{ac'}_Q + S^{ca'}_q \gamma_{\nu} \tilde{S}^{ab'}_{Q} \gamma_{\mu} S^{bc'}_Q - S^{ca'}_q \gamma_{\nu} \tilde{S}^{bb'}_{Q} \gamma_{\mu} S^{ac'}_Q - S^{cb'}_q \gamma_{\nu} \tilde{S}^{aa'}_{Q} \gamma_{\mu} S^{bc'}_Q \nonumber \\
&+& S^{cb'}_q \gamma_{\nu} \tilde{S}^{ba'}_{Q} \gamma_{\mu} S^{ac'}_Q - S^{cc'}_q Tr\Big[  S^{ba'}_Q  \gamma_{\nu} \tilde{S}^{ab'}_{Q} \gamma_{\mu} \Big] + S^{cc'}_Q Tr\Big[  S^{ba'}_Q  \gamma_{\nu} \tilde{S}^{ab'}_{q} \gamma_{\mu} \Big] \nonumber \\
&+& S^{cc'}_q Tr\Big[  S^{bb'}_Q  \gamma_{\nu} \tilde{S}^{aa'}_{Q} \gamma_{\mu} \Big] + S^{cc'}_Q Tr\Big[  S^{bb'}_Q  \gamma_{\nu} \tilde{S}^{aa'}_{q} \gamma_{\mu} \Big] - S^{cc'}_Q Tr\Big[  S^{ba'}_q  \gamma_{\nu} \tilde{S}^{ab'}_{Q} \gamma_{\mu} \Big] \nonumber \\
&+& S^{cc'}_Q Tr\Big[  S^{bb'}_q  \gamma_{\nu} \tilde{S}^{aa'}_{Q} \gamma_{\mu} \Big] \Big\}|\psi_0\rangle.
\end{eqnarray}
In  the fixed point gauge, we choose 
\begin{eqnarray}\label{lightq}
S_q^{ij}(x)&=&
\frac{i}{2\pi^2}\delta^{ij}\frac{1}{(x^2)^2}\not\!x
-\frac{m_q }{ 4\pi^2} \delta^ { ij } \frac { 1}{x^2} + \chi^i_q(x)\bar{\chi}^j_q(0) 
-\frac{ig_s}{32\pi^2}F_{\mu\nu}^A(0)t^{ij,A}\frac{1}{x^2}[\not\!x\sigma^{\mu\nu}+\sigma^{\mu\nu}\not\!x] +\cdots, \nonumber\\
\end{eqnarray}
for the light quark and 
\begin{eqnarray}\label{heavyQ}
S_Q^{ij}(x)&=&\frac{i}{(2\pi)^4}\int d^4k e^{-ik \cdot x} \left\{\frac{\delta_{ij}}{\!\not\!{k}-m_Q}
-\frac{g_sF_{\mu\nu}^A(0)t^{ij,A}}{4}\frac{\sigma_{\mu\nu}(\!\not\!{k}+m_Q)+(\!\not\!{k}+m_Q)
\sigma_{\mu\nu}}{(k^2-m_Q^2)^2}\right.\nonumber\\
&&\left.+\frac{\pi^2}{3} \langle \frac{\alpha_sGG}{\pi}\rangle\delta_{ij}m_Q \frac{k^2+m_Q\!\not\!{k}}{(k^2-m_Q^2)^4}+\cdots\right\} \, ,
\end{eqnarray}
for the heavy quark propagators. In Eq.~(\ref{lightq}), $\chi^i_q$ and $\bar{\chi}^j_q$ are the Grassmann background quark fields. In Eqs.~(\ref{lightq})  and (\ref{heavyQ}),
$F_{\mu\nu}^A$ are classical background gluon fields, and $t^{ij,A}=\frac{\lambda ^{ij,A}}{2}$ with $
\lambda ^{ij, A}$ being  the standard Gell-Mann matrices. After, replacing the above explicit forms of the propagators in the correlation function in Eqs.~(\ref{correPi1}-\ref{correPi2}), the products
of the Grassmann background quark fields and classical background gluon fields which correspond to the ground-state matrix elements of the corresponding quark and gluon operators \cite{Cohen:1994wm} are obtained,
\begin{eqnarray}\label{fields}
\chi_{a\alpha}^{q}(x)\bar{\chi}_{b\beta}^{q}(0)&=&\langle q_{a\alpha}(x)\bar{q}_{ b\beta}(0)\rangle_{\rho_N},
 \nonumber \\
F_{\kappa\lambda}^{A}F_{\mu\nu}^{B}&=&\langle
G_{\kappa\lambda}^{A}G_{\mu\nu}^{B}\rangle_{\rho_N}, \nonumber \\
\chi_{a\alpha}^{q}\bar{\chi}_{b\beta}^{q}F_{\mu\nu}^{A}&=&\langle
q_{a\alpha}\bar{q}_{ b\beta}G_{\mu\nu}^{A}\rangle_{\rho_N},
\end{eqnarray}
where, $\rho_N$ is the density of the cold nuclear matter. The matrices in the right hand sides of Eq.~(\ref{fields}) contain the in-medium quark, gluon and mixed condensates. These matrices are parametrised  as \cite{Cohen:1994wm}: i) quark condensate
\begin{eqnarray} \label{quarkfield}
&& \langle q_{a\alpha}(x)\bar{q}_{b\beta}(0)\rangle_{\rho_N}=-\frac{\delta_{ab}}{12} \Bigg[\Bigg(\langle\bar{q}q\rangle_{\rho_N}+x^{\mu}\langle\bar{q}D_{\mu}q\rangle_{
\rho_N} + \frac{1}{2}x^{\mu}x^{\nu}\langle\bar{q}D_{\mu}D_{\nu}q\rangle_{\rho_N} +...\Bigg)\delta_{\alpha\beta}
 \nonumber \\
&&+\Bigg(\langle\bar{q}\gamma_{\lambda}q\rangle_{\rho_N}+x^{\mu}\langle\bar{q}
\gamma_{\lambda}D_{\mu} q\rangle_{\rho_N}+\frac{1}{2}x^{\mu}x^{\nu}\langle\bar{q}\gamma_{\lambda}D_{\mu}D_{\nu}
q\rangle_{\rho_N}
+...\Bigg)\gamma^{\lambda}_{\alpha\beta} \Bigg],\nonumber \\
\end{eqnarray}
ii) qluon condensate
\begin{eqnarray}\label{gluonfield}
 \langle
G_{\kappa\lambda}^{A}G_{\mu\nu}^{B}\rangle_{\rho_N}&=&\frac{\delta^{AB}}{96}
\Bigg[ \langle
G^{2}\rangle_{\rho_N}(g_{\kappa\mu}g_{\lambda\nu}-g_{\kappa\nu}g_{\lambda\mu})+O(\langle \textbf{E}^{2}+\textbf{B}^{2}\rangle_{\rho_N})\Bigg],
\end{eqnarray}
where the term $O(\langle \textbf{E}^{2}+\textbf{B}^{2}\rangle_{\rho_N})$ is neglected because of its small contribution. And iii) quark-gluon mixed condensate
\begin{eqnarray} \label{mixedfield}
&&\langle g_{s}q_{a\alpha}\bar{q}_{b\beta}G_{\mu\nu}^{A}\rangle_{\rho_N}=-\frac{t_{ab}^{A }}{96}\Bigg\{\langle g_{s}\bar{q}\sigma\cdot Gq\rangle_{\rho_N}
\Bigg[\sigma_{\mu\nu}+i(u_{\mu}\gamma_{\nu}-u_{\nu}\gamma_{\mu}) \!\not\! {u}\Bigg]_{\alpha\beta} +\langle g_{s}\bar{q}\!\not\! {u}\sigma\cdot Gq\rangle_{\rho_N} \Bigg[\sigma_{\mu\nu}\!\not\!
{u} \nonumber \\
&&+i(u_{\mu}\gamma_{\nu}-u_{\nu}\gamma_{\mu} )\Bigg]_{\alpha\beta}-4\Bigg(\langle\bar{q}u\cdot D u\cdot D
q\rangle_{\rho_N}
+im_{q}\langle\bar{q}
\!\not\! {u}u\cdot D q\rangle_{\rho_N}\Bigg) \Bigg[\sigma_{\mu\nu}+2i(u_{\mu}\gamma_{\nu}-u_{\nu}\gamma_{\mu}
)\!\not\! {u}\Bigg]_{\alpha\beta}\Bigg\},
\nonumber \\
\end{eqnarray}
where $D_\mu=\frac{1}{2}(\gamma_\mu \!\not\!{D}+\!\not\!{D}\gamma_\mu)$. The in-medium modification of the different condensates in Eqs.~(\ref{quarkfield}-\ref{mixedfield}) are defined as follows:
\begin{eqnarray} \label{ }
\langle\bar{q}\gamma_{\mu}q\rangle_{\rho_N}&=&\langle\bar{q}\!\not\!{u}q\rangle_{\rho_N} u_{\mu} ,\nonumber \\
\langle\bar{q}D_{\mu}q\rangle_{\rho_N}&=&\langle\bar{q}u\cdot D q\rangle_{\rho_N} u_{\mu}=-im_{q}\langle\bar{q}\!\not\!{u}q\rangle_{\rho_N} u_{\mu}  ,\nonumber \\
\langle\bar{q}\gamma_{\mu}D_{\nu}q\rangle_{\rho_N}&=&\frac{4}{3}\langle\bar{q} \!\not\! {u}u\cdot D
q\rangle_{\rho_N}(u_{\mu}u_{\nu}-\frac{1}{4}g_{\mu\nu}) +\frac{i}{3}m_{q}
\langle\bar{q}q\rangle_{\rho_N}(u_{\mu}u_{\nu}-g_{\mu\nu}),
\nonumber \\
\langle\bar{q}D_{\mu}D_{\nu}q\rangle_{\rho_N}&=&\frac{4}{3}\langle\bar{q}
u\cdot D u\cdot D
q\rangle_{\rho_N}(u_{\mu}u_{\nu}-\frac{1}{4}g_{\mu\nu}) -\frac{1}{6} \langle
g_{s}\bar{q}\sigma\cdot Gq\rangle_{\rho_N}(u_{\mu}u_{\nu}-g_{\mu\nu}) , \nonumber \\
\langle\bar{q}\gamma_{\lambda}D_{\mu}D_{\nu}q\rangle_{\rho_N}&=&2\langle\bar{q}
\!\not\! {u}u\cdot D u\cdot D q\rangle_{\rho_N}
\Bigg[u_{\lambda}u_{\mu}u_{\nu} -\frac{1}{6}
(u_{\lambda}g_{\mu\nu}+u_{\mu}g_{\lambda\nu}+u_{\nu}g_{\lambda\mu})\Bigg]\nonumber\\
&&-\frac{1}{6} \langle g_{s}\bar{q}\!\not\! {u}\sigma\cdot
Gq\rangle_{\rho_N}(u_{\lambda}u_{\mu}u_{\nu}-u_{\lambda}g_{\mu\nu}).\nonumber
\\
\end{eqnarray}

On the basis of Lorentz covariance, parity, and time reversal considerations, the QCD side of the correlation function in nuclear matter can be decomposed over the Lorentz structures as follows:
\begin{eqnarray}\label{ }
\Pi^{QCD}_{\mu\nu} (p) &=& \Pi^{QCD}_1(p^2, p_0) \, g_{\mu\nu} + \Pi^{QCD}_2(p^2, p_0) \, g_{\mu\nu} \, \pslash +  \Pi^{QCD}_3(p^2, p_0) \, g_{\mu\nu} \, \uslash + \Pi^{QCD}_4(p^2, p_0) \, u_{\mu}u_{\nu} \nonumber \\
&+& \Pi^{QCD}_5(p^2, p_0) \, u_{\mu}u_{\nu} \, \pslash  + \Pi^{QCD}_6(p^2, p_0) \, u_{\mu}u_{\nu} \, \uslash,
\end{eqnarray}
nevertheless, in the vacuum limit the coefficients, $\Pi^{QCD}_{i=3,...,6}(p^2, p_0)$,  are vanished. As is seen from the existing structures in above equation, we have applied the same procedure in QCD side as the physical one to remove the spin$-1/2$ pollution.   The necessary in-medium QCD sum rules for the physical parameters of the doubly heavy baryons can be obtained by equating the coefficients of the same structures in both $\Pi_{\mu\nu}^{Phys}(p)$ and $\Pi_{\mu\nu}^{QCD}(p)$ functions. In the nuclear matter, the invariant amplitudes $\Pi^{QCD}_{i}(p^2, p_0)$ corresponding to each structure can be represented as the dispersion integral,
\begin{equation}
\label{ }
\Pi^{QCD}_{i} (p^2, p_0) = \int_{(m_Q+m_{Q'})^2}^{\infty}\frac{\rho_i^{QCD}(s, p_0)}{s-p^2}ds,
\end{equation}  
where $\rho_i^{QCD}(s, p_0)$ is the two-point spectral density for $i^{th}$ structure. As the standard procedure, $\rho_i^{QCD}(s, p_0)$ is obtained from the imaginary part of the correlation function. The technical methods used in the calculations of the components  of the spectral densities are detailed in Ref.~\cite{Azizi:2018dtb}. In the correlation functions $\Pi^{QCD}_{i} (p^2, p_0)$, the spectral densities $\rho_i^{QCD}(s, p_0)$ for $i=g_{\mu\nu}, g_{\mu\nu} \, \pslash$ and $g_{\mu\nu} \, \uslash$ are given in the appendix.

After the Borel transformation on the variable $p^2$ and performing the continuum subtraction,  we get,
\begin{equation}
\label{ }
\Pi^{QCD}_{i} (M^2, s_0^*, p_0) = \int_{(m_Q+m_{Q'})^2}^{s_0^*}ds \rho_i^{QCD}(s, p_0)e^{\frac{-s}{M^2}},
\end{equation}  
where $s_0^*$ is the in-medium continuum threshold. After matching the coefficients of different structures derived from the physical and QCD sides of the correlation function, we get the following sum rules to be applied in the calculations of mass, residue and vector self-energy of the spin-$3/2$ doubly heavy baryons:
 \begin{eqnarray}\label{sumrules}
m^{*}_{DH} \lambda^{*2}_{DH} e^{-\mu^2/M^2}& = & \Pi^{QCD}_1(M^2, s_0^*, p_0), \nonumber  \\
 \lambda^{*2}_{DH} e^{-\mu^2/M^2}& = &\Pi^{QCD}_2(M^2, s_0^*, p_0), \nonumber  \\
\Sigma_\upsilon \lambda^{*2}_{DH} e^{-\mu^2/M^2}& = &\Pi^{QCD}_3(M^2, s_0^*, p_0). 
\end{eqnarray}
These coupled sum rules will be simultaneously solved to find the physical quantities under consideration.

\section{The Analysis of the sum rules}
In this section, the sum rules in  Eq.~(\ref{sumrules}) are used to obtain the numerical values of the vacuum and in-medium mass and residue of the heavy $\Xi^{*}_{QQ'}$ and $\Omega^{*}_{QQ'}$ baryons and their in-medium vector and scalar self energies. These sum rules contain different parameters:  the mass of light and heavy quarks as well as the vacuum and in-medium quark, gluon and mixed condensates with different dimensions. Collected from different sources \cite{Cohen:1994wm,PhysRevC.47.2882,PhysRevD.98.030001}, we present their numerical values as: $\rho_N^{sat}=(0.11)^3 ~GeV^3$, $ \langle q^{\dag} q\rangle_{\rho_N}=\frac{3}{2}\rho_N$,  $ \langle s^{\dag} s\rangle_{\rho_N}=0$, $ \langle \bar{q}q\rangle_{0}=(-0.241)^3~GeV^3$,  $\langle \bar{s}s\rangle_{0} = 0.8 \langle \bar{q}q\rangle_{0}$,  $\langle \bar{q}q\rangle_{\rho_N}=\langle \bar{q}q\rangle_{0}+\frac{\sigma_N}{2 m_q}\rho_N$, $\sigma_N=0.059~GeV$, $m_q=0.00345~GeV$, $\langle \bar{s}s\rangle_{\rho_N}=\langle \bar{s}s\rangle_{0}+y\frac{\sigma_N}{2 m_q}\rho_N$, $y=0.05\pm0.01$, $\langle \frac{\alpha_s}{\pi}G^2\rangle_{0}=(0.33\pm0.04)^4~GeV^4$, $\langle \frac{\alpha_s}{\pi}G^2\rangle_{\rho_N}=\langle \frac{\alpha_s}{\pi}G^2\rangle_{0}-(0.65\pm 0.15)~GeV~\rho_N$, $\langle q^{\dag}iD_0 q\rangle_{\rho_N}=0.18~GeV~\rho_N$, $\langle s^{\dag}iD_0 s\rangle_{\rho_N}=\frac{m_s\langle \bar{s}s\rangle_{\rho_N}}{4}+0.02~GeV~\rho_N$, $\langle \bar{q}iD_0q\rangle_{\rho_N}=\langle \bar{s}iD_0s\rangle_{\rho_N}=0$, $\langle \bar{q}g_s\sigma G q\rangle_{0}=m_0^2 \langle \bar{q}q\rangle_{0}$, $\langle \bar{s}g_s\sigma G s\rangle_{0}=m_0^2 ~\langle \bar{s}s\rangle_{0}$, $m_0^2=0.8 ~GeV^2$, $\langle \bar{q}g_s\sigma G q\rangle_{\rho_N}=\langle \bar{q}g_s\sigma G q\rangle_{0}+ 3~GeV^2~\rho_N$, $\langle \bar{s}g_s\sigma G s\rangle_{\rho_N}=\langle \bar{s}g_s\sigma G s\rangle_{0}+ 3y~GeV^2~\rho_N$, $\langle q^{\dag}g_s\sigma G q\rangle_{\rho_N}=-0.33 ~GeV^2~\rho_N$, $\langle q^{\dag}iD_0 iD_0 q\rangle_{\rho_N}=0.031~GeV^2~\rho_N-\frac{1}{12}\langle q^{\dag}g_s\sigma G q\rangle_{\rho_N}$, $\langle s^{\dag}g_s\sigma G s\rangle_{\rho_N}=-0.33y ~GeV^2~\rho_N$, $\langle s^{\dag}iD_0 iD_0 s\rangle_{\rho_N}=0.031y~GeV^2~\rho_N-\frac{1}{12}\langle s^{\dag}g_s\sigma G s\rangle_{\rho_N}$, $m_u=2.2_{-0.4}^{+0.5}~MeV$, $m_d=4.7_{-0.3}^{+0.5}~MeV$, $m_s=0.13~GeV$, $m_b=4.78\pm0.06~GeV$, $m_c= 1.67\pm 0.07GeV$. Note that each condensate at dense medium can be written up to the first order in nucleon density as $\langle\hat{O}\rangle_{\rho_N}=\langle\hat{O}\rangle_{0} +\rho_N\langle\hat{O}\rangle_{N}$, where $\langle\hat{O}\rangle_{0}$ is its vacuum value and $\langle\hat{O}\rangle_{N}$ is its value between one-nucleon states.

The sum rules in Eq.~(\ref{sumrules}) contain two auxiliary parameters in addition to the input parameters represented above: the continuum threshold $s_0^*$ and the Borel mass parameter $M^2$. For the quality of the numerical results of the physical quantities, we should minimize the dependence of the results on these parameters. To this end, we require the pole dominance and impose the condition:
\begin{equation}
\label{ }
\text{PC}=\frac{\int^{s_0^*}_{(m_Q+m'_Q)^2}ds \rho(s, p_0)e^{-\frac{s}{M^2}}}{\int_{(m_Q+m'_Q)^2}^{\infty}ds \rho(s, p_0)e^{-\frac{s}{M^2}}} \geqslant \frac{1}{2}.
\end{equation}
Following these conditions, we choose the in-medium threshold in the interval $s_0^*=(m_{DH}+[0.5-0.7])^2 ~GeV^2$. 

The next step is to fix  the Borel mass parameter. For this aim, we consider again the pole dominance and convergence of the series obtained in the QCD side of the correlation function. Thus, the upper limit of this parameter is obtained considering the dominance of pole contribution over the contribution of the higher states and continuum and its lower limit is found requiring the convergence of the series of different operators and imposing the condition of exceeding the perturbative part over the total non-perturbative contributions. Following these criteria, we obtain:  
\begin{equation}
\label{ }
[M^2_{\textrm{min}},M^2_{\textrm{max}}]=\left\{\begin{array}{c} $[3-5 ]$ ~ GeV^2 ~~ \textrm{for} ~\Xi^{*}_{cc} ~\textrm{and}  ~\Omega^{*}_{cc}, \\ 
$[6-8 ]$ ~ GeV^2 ~~ \textrm{for} ~\Xi^{*}_{bc} ~\textrm{and}  ~\Omega^{*}_{bc}, \\
$[8-12 ]$ ~ GeV^2 ~~ \textrm{for} ~\Xi^{*}_{bb} ~\textrm{and}  ~\Omega^{*}_{bb}.
\end{array}\right.
\end{equation}

In Fig.1, we show the pole contribution, for instance at $\Xi_{cc}^*$ channel, in terms of Borel parameter and at three fixed values of continuum threshold in their working interval. In average, we find  PC$=0.71$, which ensures the pole dominance of the related  channel for the structure $g_{\mu\nu} \, \pslash$.
\begin{figure}\label{PoleCont}
\centering
\includegraphics[width=0.6\textwidth]{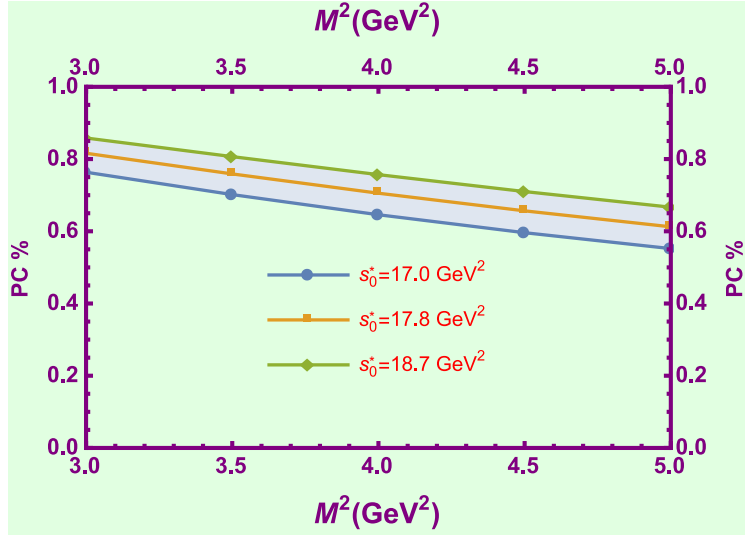}
\caption{ The pole contribution in the $\Xi_{cc}^*$ channel as a function of Borel mass $M^2$ at saturation density and at fixed values of the in-medium continuum threshold.}  
\end{figure}

Before analyses of the results in terms of the medium density, we would like to turn off the density and present the results in vacuum. Obtained from the analyses, our predictions for the masses of the considered states in vacuum are presented in table II. In the literature, there are a plenty of studies on  the vacuum mass of the spin$-3/2$ doubly heavy baryons listed in the same table. We refer to some of them, which are extracted using: an Extended Choromomagnetic Model (ECM) \cite{PhysRevD.97.054008}, the Bethe Salpeter Equation Approach (BSEA) \cite{Yu:2018com}, Vacuum QCD Sum Rule (VQCDSR) \cite{PhysRevD.78.094007,0954-3899-40-6-065003} , a Relativized Quark Model (RQM) \cite{PhysRevD.66.014008,PhysRevD.96.114006}, the Feynman-Hellmann Theorem (FHT) \cite{PhysRevD.52.1722} , Quark Model (QM) \cite{Roberts:2007ni}, Lattice QCD (LQCD)  \cite{PhysRevD.90.094507} and a Bag Model (BM) \cite{PhysRevD.70.094004}.
\begin{table}[h]
\centering
\begin{tabular}{|lcccccc|}\hline 
&$\Xi_{cc}^*$&$\Omega_{cc}^*$&$\Xi_{bc}^*$&$\Omega_{bc}^*$&$\Xi_{bb}^*$&$\Omega_{bb}^*$ \\
PW ($\rho_N \rightarrow 0$) &$3.73\pm0.07$&$3.76\pm0.05$&$6.98\pm0.09$&$7.05\pm0.07$&$10.06\pm0.12$&$10.11\pm0.10$ \\
ECM \cite{PhysRevD.97.054008} &$3.696\pm0.0074$ & $3.802\pm0.008$ & $6.973\pm0.0055$ & $7.065\pm0.0075$ & $10.188 \pm0.0071$ & $10.267\pm0.012$ \\
BSEA\cite{Yu:2018com} & $3.62\pm0.01$ & $3.71\pm0.02$ & $6.99\pm0.02$ & $7.07\pm0.01$ & $10.27\pm0.01$ & $10.35 \pm0.01$ \\
VQCDSR\cite{0954-3899-40-6-065003} &$3.69\pm0.16 $ &  $3.78\pm0.16$ & $7.25\pm0.20$ & $7.3\pm0.2$ & $10.4\pm1.0$ & $10.5\pm0.2 $ \\
RQM\cite{PhysRevD.96.114006} & $3.675$ & $3.772$ & - & - & $10.169$ & $10.258$ \\
RQM\cite{PhysRevD.66.014008} & $3.727$ & $3.872$ & $6.98$ & $ 7.13$ & $10.237$ & $10.389$ \\
FHT\cite{PhysRevD.52.1722} & $3.74\pm0.07$ & $3.82\pm0.08$ & $7.06\pm0.09$ & $7.12\pm0.09$ & $10.37\pm0.1$ & $10.40\pm0.1$ \\
QM \cite{Roberts:2007ni}  & $3.753$ & $3.876$ & $7.074$ & $7.187$ & $10.367$ & $10.486$ \\
LQCD \cite{PhysRevD.90.094507} & $3.692(28)(21)$ & $3.822(20)(22)$ & $6.985(36)(28)$ & $7.059(28)(21)$ & $10.178(30)(24)$ & $10.308(27)(21)$ \\
BM\cite{PhysRevD.70.094004} & $3.59$ & $3.77$ & $6.85$ & $7.02$ & $10.11$ & $10.29$ \\
VQCDSR\cite{PhysRevD.78.094007} & $3.90\pm0.10$ & $3.81 \pm0.06$ & $8.0\pm0.26$ & $7.54\pm0.08$ & $10.35\pm0.08$ & $10.28\pm0.05$
\\\hline
\end{tabular}
\caption{Vacuum mass of the spin$-3/2$ doubly heavy baryons compared with the literature. The units are in GeV and PW ($\rho_N \rightarrow 0$) means the present work at $\rho_N=0$.}
\end{table}
Looking at this table, we see that the results from different approaches are over all consistence with each other within the errors. 
These results can be verified in future experiments.   
\begin{figure}[h]
\centering
\subfloat[]{\includegraphics[width=0.45\textwidth]{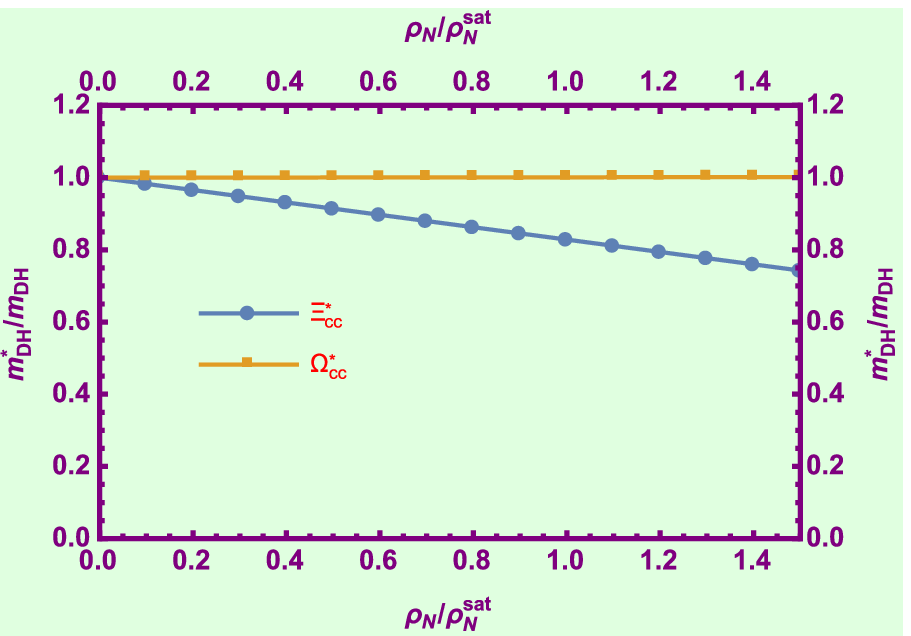}}~~~~~~~~
\subfloat[]{ \includegraphics[width=0.45\textwidth]{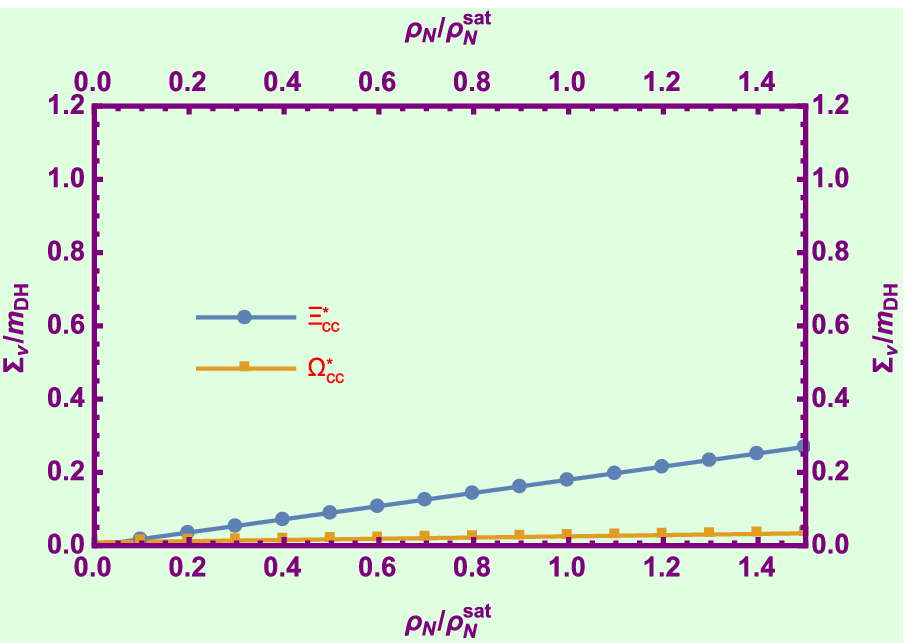}}\\
\vspace{0.5cm}
\subfloat[]{\includegraphics[width=0.45\textwidth]{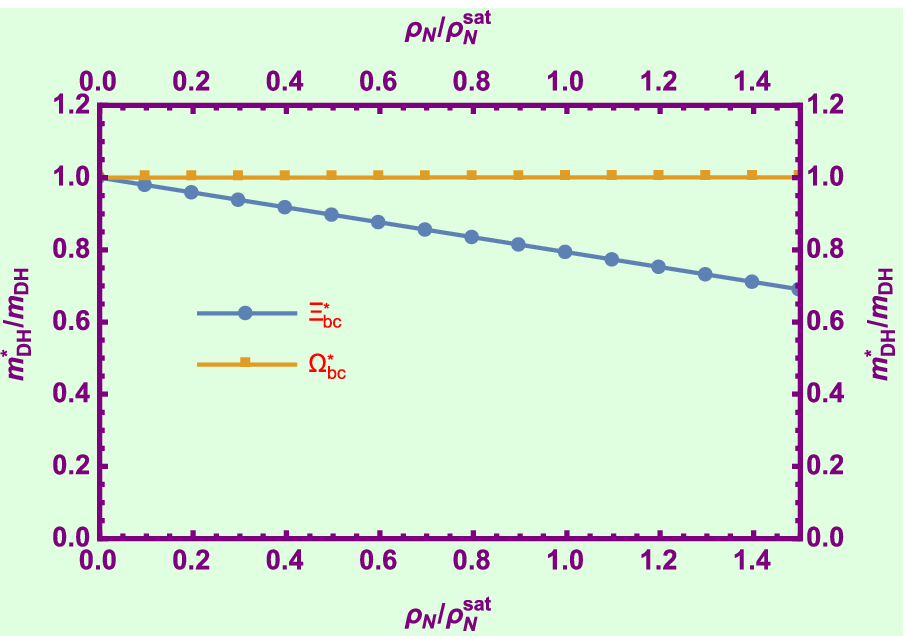}}~~~~~~~~
\subfloat[]{ \includegraphics[width=0.45\textwidth]{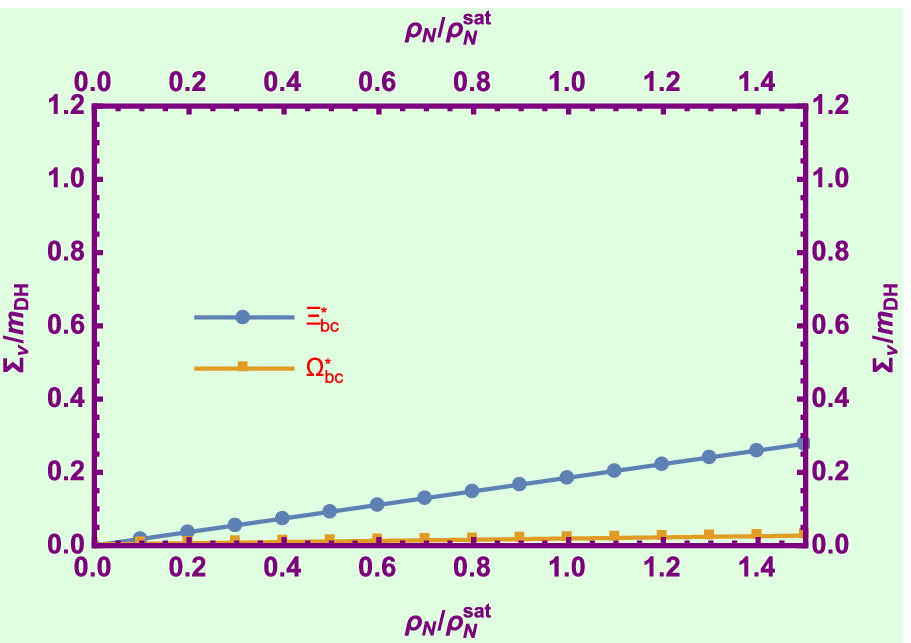}}\\
\vspace{0.5cm}
\subfloat[]{\includegraphics[width=0.45\textwidth]{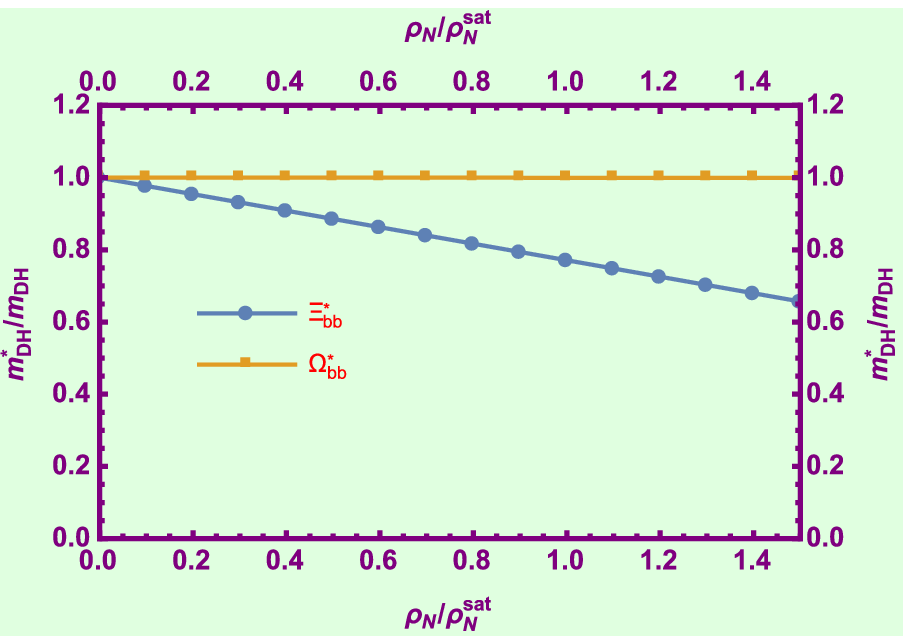}}~~~~~~~~
\subfloat[]{ \includegraphics[width=0.45\textwidth]{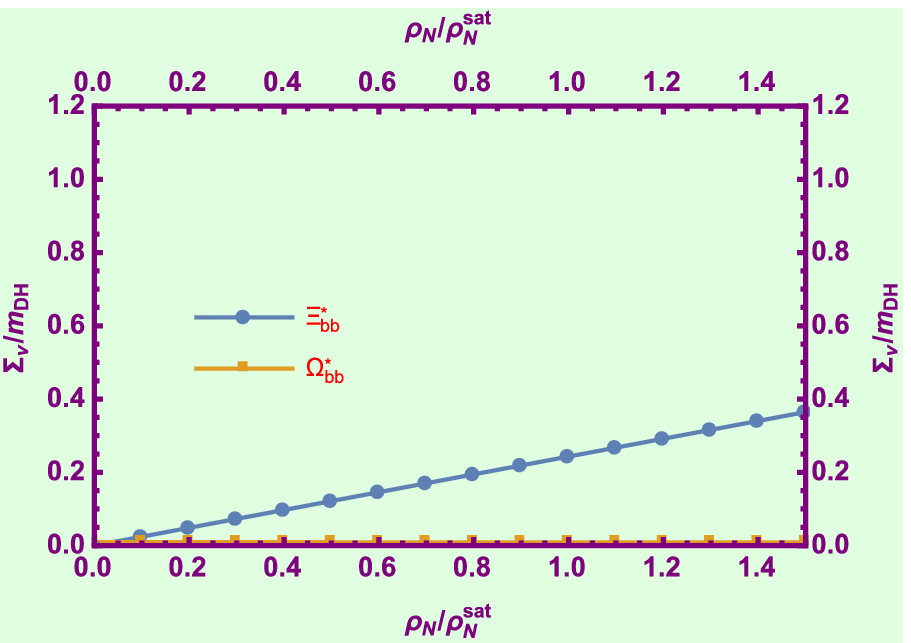}}
\caption{The in-medium mass to vacuum mass ratio, $m_{DH}^*/m_{DH}$, (left panel) and the vector self-energy to vacuum mass ratio, $\Sigma_{\nu}/m_{DH}$, (right panel) with respect to $\rho_N/\rho^{sat}_N$ for the doubly heavy  $\Xi^{*}_{QQ'}$ and $\Omega^{*}_{QQ'}$ baryons at average values of the continuum threshold and Borel mass parameter.}
\end{figure}

\begin{figure}[h]
\centering
\subfloat[]{\includegraphics[width=0.45\textwidth]{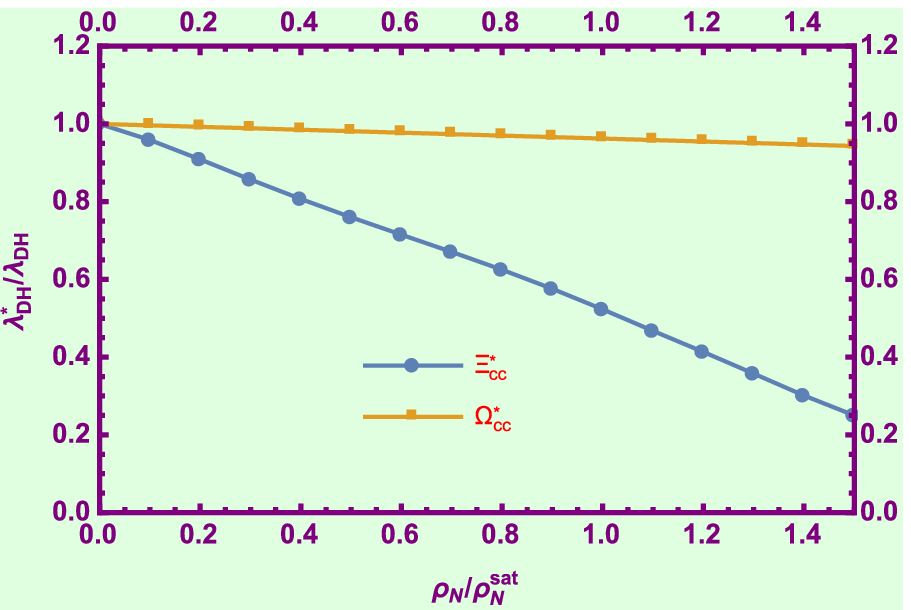}}~~~~~~~~
\subfloat[]{ \includegraphics[width=0.45\textwidth]{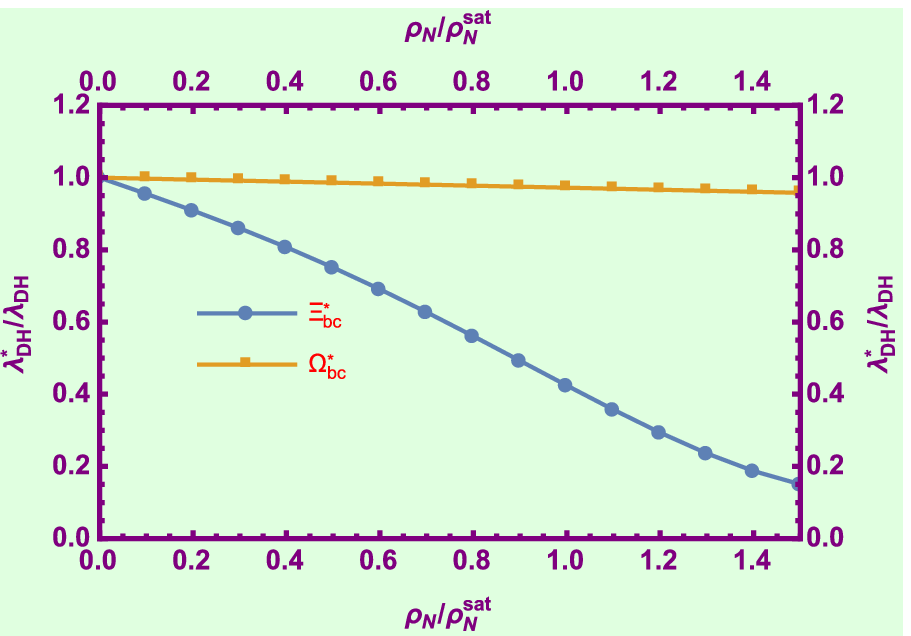}}\\
\subfloat[]{\includegraphics[width=0.45\textwidth]{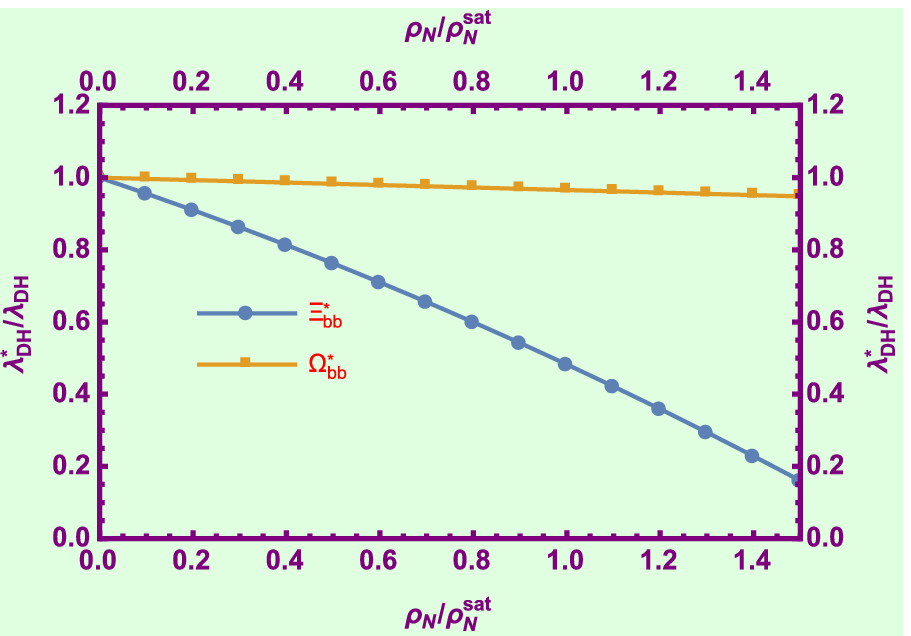}}
\caption{The ratio  $\lambda_{DH}^*/\lambda_{DH}$ as  a function of $\rho_N / \rho^{sat}_N$ for the doubly heavy $\Xi^{*}_{QQ'}$ and $\Omega^{*}_{QQ'}$ baryons at average values of the continuum threshold and Borel mass parameter.}  
\end{figure}
In  further analyses, we discuss the density dependence of the results. Thus, in Figs. 2 and 3, we plot the ratios  $m_{DH}^*/m_{DH}$, $\Sigma_{\nu}/m_{DH}$ and  $\lambda_{DH}^*/\lambda_{DH}$ as  functions of $\rho_N / \rho^{sat}_N$ at average values of the continuum threshold and $M^2$ for different members of the doubly heavy baryons. In the calculations, we use the $\rho^{sat}_N=(0.11)^3 ~GeV^3$ for the saturation density of the medium. This is equivalent to roughly $2.5\times10^{14}$ g/cm$^3$ which corresponds to $\sim 20\%$ of the density of the neutron stars' core. From these figures, we see that the baryons $\Omega^{*}_{cc}$,  $\Omega^{*}_{bc}$ and $\Omega^{*}_{bb}$ do not see the medium at all. This can be attributed to the fact that the quark contents of these baryons, i.e. $ccs$, $cbs$ and $bbs$ are different than the quark content of the medium which is considered to be $uud/udd$ and the strange and charm components of the nucleons are ignored. The baryons $\Xi^{*}_{cc}$,  $\Xi^{*}_{bc}$ and $\Xi^{*}_{bb}$, however, interact with the medium due to their $u/d$ quark contents. The mass of these baryons show strong dependence on the density of the medium. Such that their masses reduce to their $74\%$, $69\%$, $66\%$ of vacuum mass values for $\Xi^{*}_{cc}$,  $\Xi^{*}_{bc}$ and $\Xi^{*}_{bb}$, respectively at $\rho_N=1.5 \rho^{sat}_N$. As is seen, the dependence of the masses of these baryons show linear dependence on the density of the medium and the shifts are negative referring to the attraction of these baryons by the medium.  
\begin{figure}[h]
\centering
\subfloat[]{\includegraphics[width=0.45\textwidth]{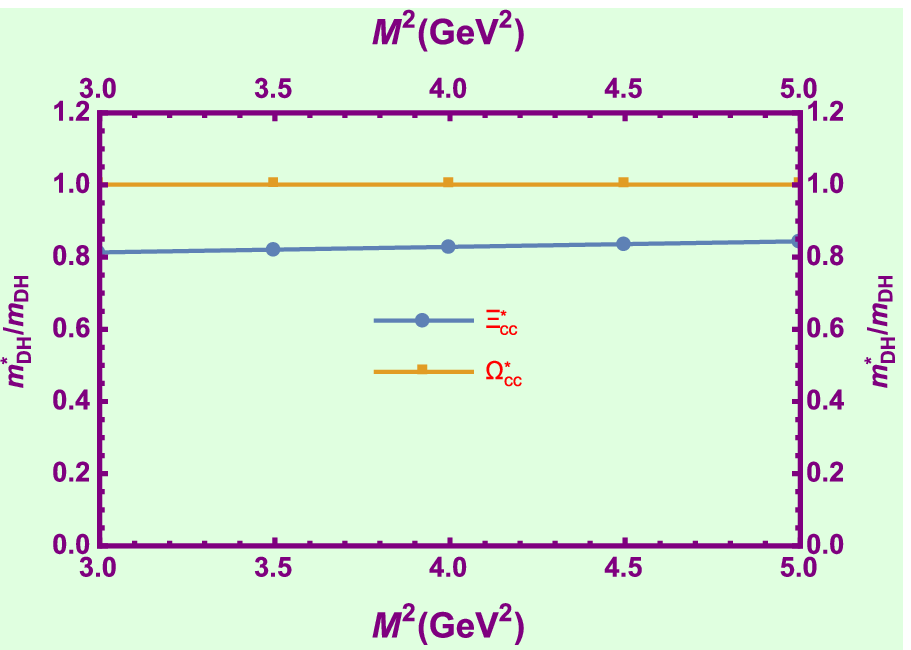}}~~~~~~~~
\subfloat[]{ \includegraphics[width=0.45\textwidth]{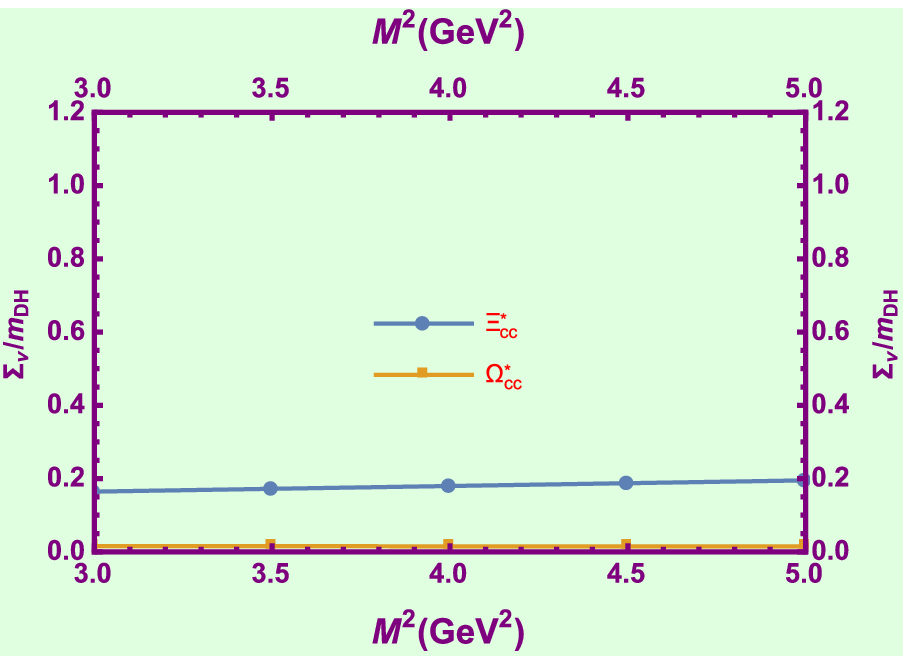}}\\
\vspace{0.5cm}
\subfloat[]{\includegraphics[width=0.45\textwidth]{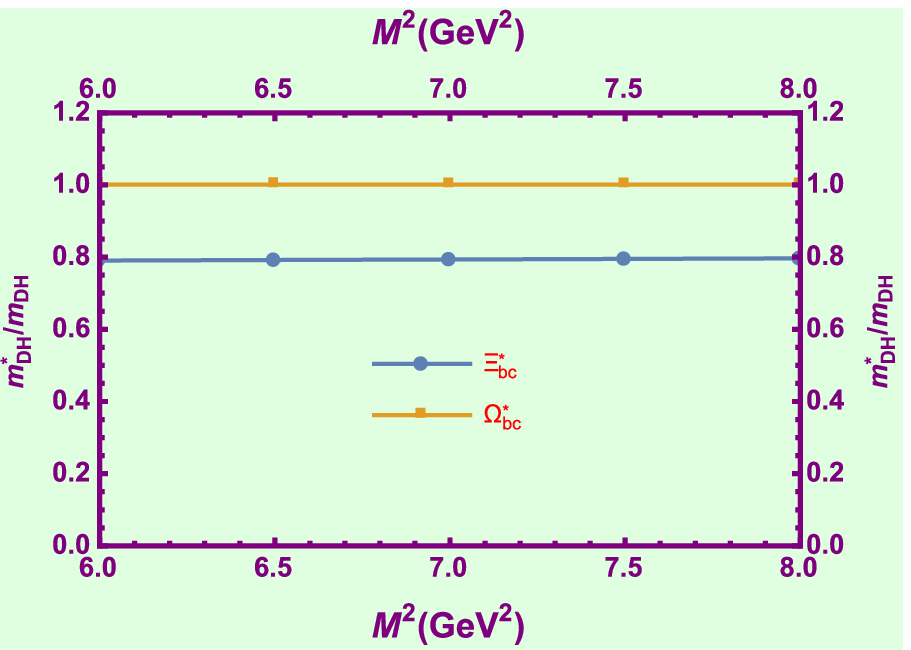}}~~~~~~~~
\subfloat[]{ \includegraphics[width=0.45\textwidth]{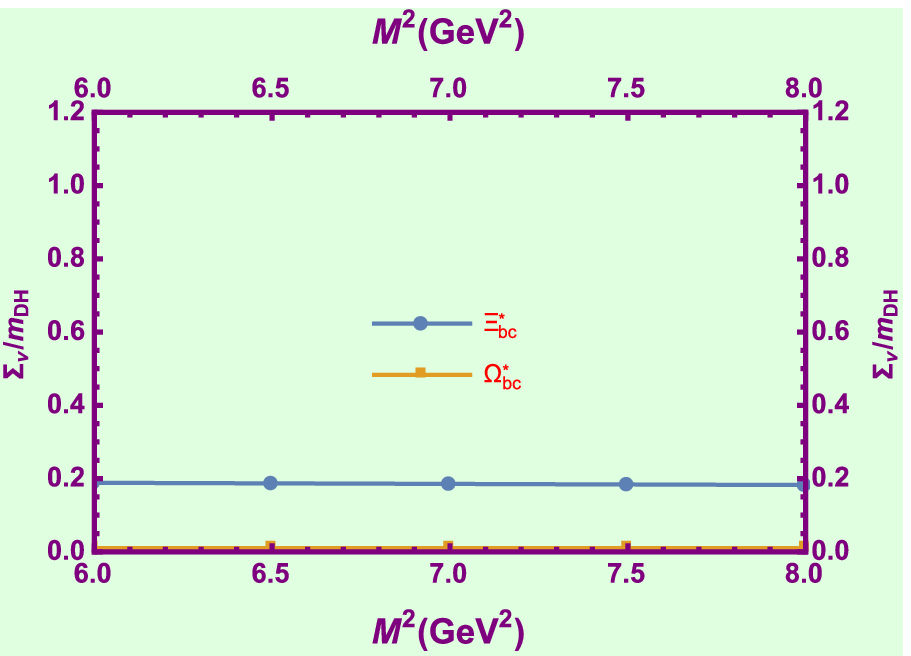}}\\
\vspace{0.5cm}
\subfloat[]{\includegraphics[width=0.45\textwidth]{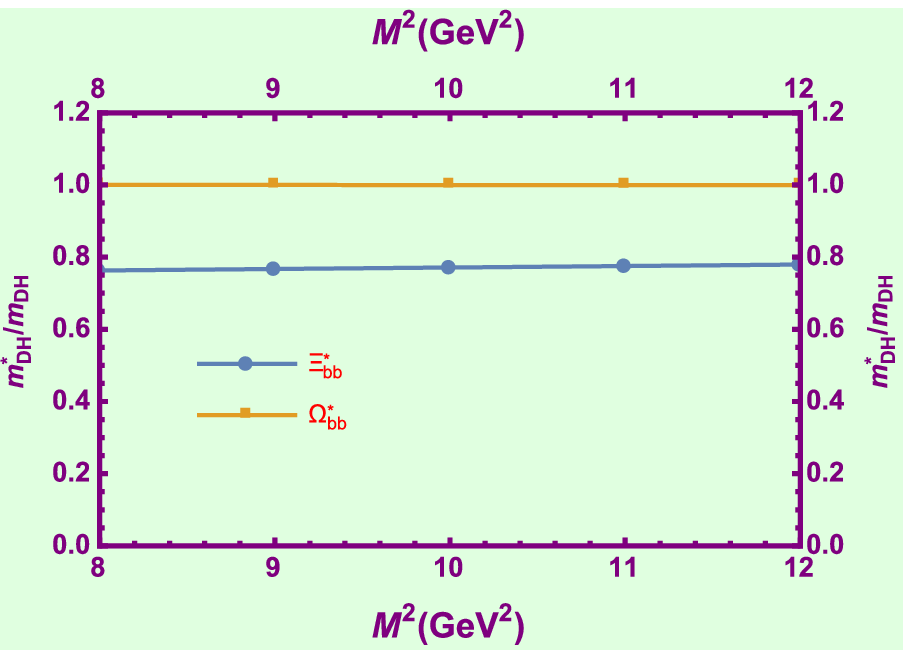}}~~~~~~~~
\subfloat[]{ \includegraphics[width=0.45\textwidth]{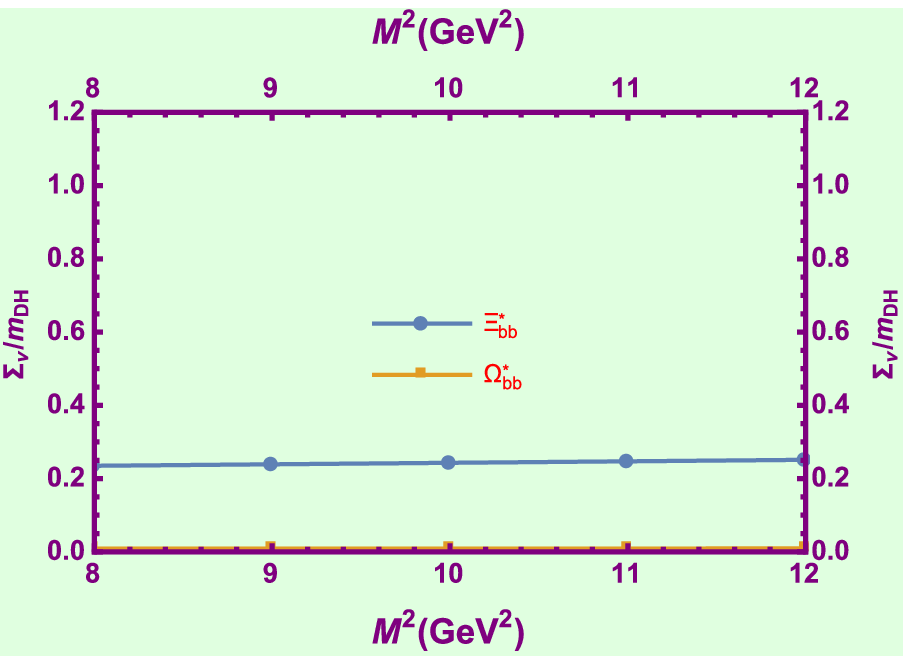}}
\caption{The dependence of the same ratios as Fig. 2 on the Borel mass parameter at the saturated nuclear matter density and average values of the continuum threshold.}  
\end{figure}
 
\begin{figure}[h]
\centering
\subfloat[]{\includegraphics[width=0.45\textwidth]{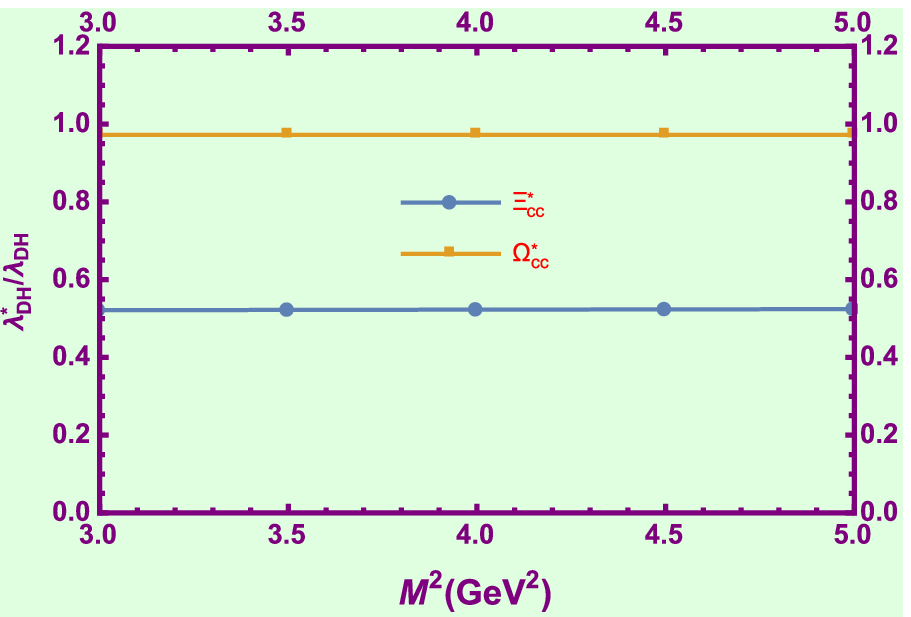}}~~~~~~~~
\subfloat[]{ \includegraphics[width=0.45\textwidth]{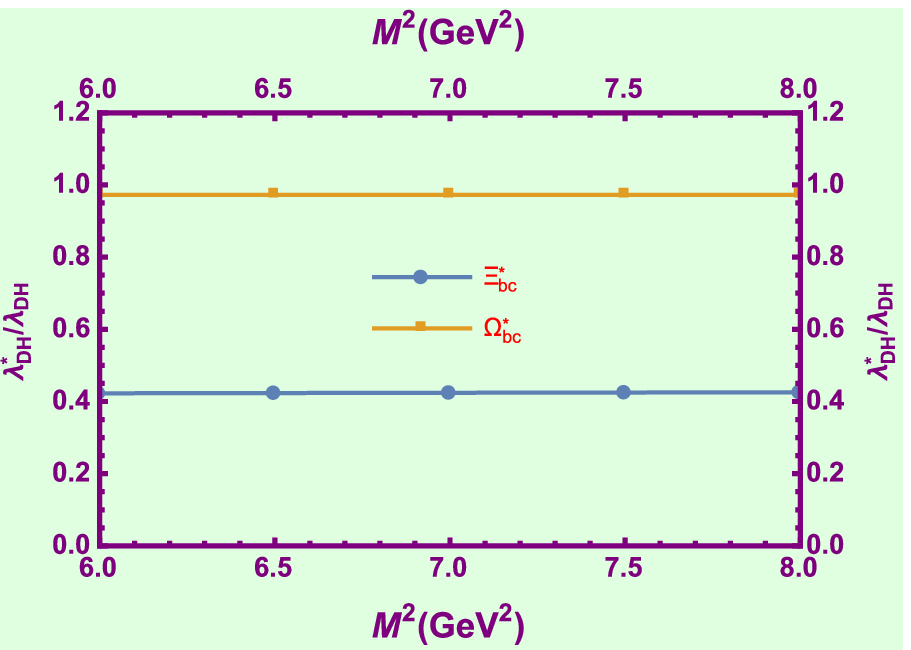}}\\
\subfloat[]{\includegraphics[width=0.45\textwidth]{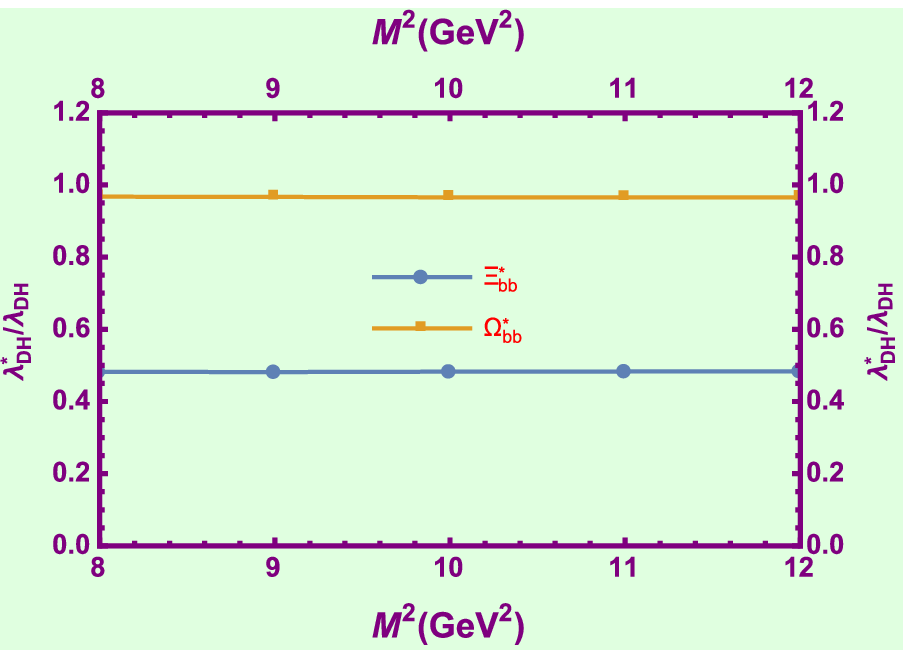}}
\caption{The dependence of the same ratio as Fig. 3 to the Borel mass parameter at the saturated nuclear matter density and average values of the continuum threshold. }  
\end{figure}
Their residues or coupling strengths show very strong sensitivity to the nuclear matter density such that, at $\rho_N=1.5 \rho^{sat}_N$ they approach  roughly $20\%$ of their vacuum values.  The dependence of the residues of these baryons to the density is roughly linear, as well. This is similar to the behavior of the nucleons spectroscopic parameters with respect to the density which are obtained to be roughly linear \cite{Azizi2014}. The vector self energies of the $\Omega^{*}_{QQ'}$ remain roughly zero by increasing the density of the medium. However, this quantity for the  $\Xi^{*}_{QQ'}$ baryons grows from zero to  approximately $27\%$ (in $\Xi^{*}_{cc}$ and  $\Xi^{*}_{bc}$ channels) and $36\%$ ( in $\Xi^{*}_{bb}$ channel)  of their vacuum mass values. We shall again note that the dependence of the  $\Sigma_{\nu}/m_{DH}$ for the $\Xi^{*}_{QQ'}$ baryons on the density of the medium is roughly linear. 

In Figs. 4 and 5, we depict the variations of the ratios,   $m_{DH}^*/m_{DH}$, $\Sigma_{\nu}/m_{DH}$ and $\lambda_{DH}^*/\lambda_{DH}$  with respect to $M^2$ at saturated nuclear matter density and average values of the continuum threshold. These figures show also that the $\Omega$ baryons do not show any response to the medium, however $\Xi$ baryons interact with the medium considerably. At nuclear matter density the ratios  $m_{DH}^*/m_{DH}$ are found to be 1 for  $\Omega$ baryons and $\sim83\%$, $\sim79\%$ and $\sim77\%$ for the baryons $\Xi^{*}_{cc}$,  $\Xi^{*}_{bc}$ and $\Xi^{*}_{bb}$, respectively. As is seen, $\Sigma_{\nu}$ is equal to zero for the $\Omega$ baryons, but $\Sigma_{\nu}/m_{DH}$ ratio for $\Xi^{*}_{cc}$,  $\Xi^{*}_{bc}$ and $\Xi^{*}_{bb}$,   are obtained as $18\%$,  $19\%$ and $24\%$ at saturated density, respectively. 
\begin{table}[h]
	\addtolength{\tabcolsep}{10pt}
	\begin{center}\begin{tabular}{lcccccc}\hline \hline 
&$\Xi_{cc}^*$&$\Omega_{cc}^*$&$\Xi_{bc}^*$&$\Omega_{bc}^*$&$\Xi_{bb}^*$&$\Omega_{bb}^*$  \\
\hline\hline
$\Sigma_S/m_{DH} $ &$-0.17$ &$\sim 0$ & $-0.21$ &$\sim 0$ &$-0.23$ &$\sim 0$ \\
$\Delta \lambda_{DH}/\lambda_{DH}$  &$-0.48$ &$-0.04$ &$-0.57$ & $-0.03$& $-0.52$&$-0.04$ \\
$\Sigma_{\nu}/m_{DH}$ & $+0.18$& $+0.03$ & $+0.19$ & $+0.02$& $+0.24$&$+0.01$ \\
\hline\hline
\end{tabular}
\end{center}
\caption{The ratios  $\Sigma_S/m_{DH} $, $\Delta \lambda_{DH}/\lambda_{DH}$ and $\Sigma_{\nu}/m_{DH}$ for   the $\Xi^{*}_{QQ'}$ and $\Omega^{*}_{QQ'}$ baryons  at average values of the continuum threshold and Borel mass parameter and at saturation nuclear matter density.}
	\label{table}
\end{table}
The variations of the considered ratios are very mild against the variations of the Borel mass parameter at saturation density and average values of the continuum threshold. We collect, the average values of these ratios at saturation nuclear matter density as well as the average values of the Borel mass square and continuum threshold in table III. From this table, we report that, the masses of the $\Omega$ baryons do not show any shifts in the medium, but their residues show very small negative shifts due to the medium. The ratio $\Sigma_{\nu}/m_{DH}$ for the $\Omega$ baryons takes very small but positive values. The scalar self energies or shifts on the masses of the $\Xi^{*}_{QQ'}$ baryons show considerable negative  values. The negative sign shows the scalar attraction of these baryons by the medium. The maximum shift belongs to the $\Xi^{*}_{bb}$ baryon but the minimum shift corresponds to the $\Xi^{*}_{cc}$ state. The negative shifts on the residues are ,$-48\%$, $-57\%$ and $-52\%$ of vacuum residues for the $\Xi^{*}_{cc}$,  $\Xi^{*}_{bc}$ and $\Xi^{*}_{bb}$, respectively. We see the positive $18\%$, $19\%$ and $24\%$ of the vacuum mass values for the vector self energies of the $\Xi^{*}_{cc}$,  $\Xi^{*}_{bc}$ and $\Xi^{*}_{bb}$ baryons, respectively, referring  to considerable vector repulsion of these states by the medium at saturated nuclear matter density and at average values of the auxiliary parameters $M^2$ and $s_0^*$. 

 \section{discussion and outlook}
The doubly heavy baryons have received  special attentions after the discovery of the $\Xi_{cc}^{++}$ baryon with double charmed quark by the LHCb Collaboration. Although its measured mass is different than that of the previous SELEX Collaboration and there is a conflict in this regard, our hope has been increased for the discovery of the other members of the doubly heavy charmed baryons, as well as the doubly heavy bottom baryons. We hope, by developing experimental studies, we will see more discoveries of the doubly heavy baryons predicted by the quark model in near future. Thus, investigation of different aspects of these particles especially their spectroscopic parameters in vacuum and in medium is of great importance. Their vacuum properties have already been discussed using various methods and approaches. In the present study, we investigate the spectroscopic properties of the doubly heavy spin$-3/2$ baryons at cold nuclear matter using the in-medium  sum rule approach. It is observed that the parameters of the $\Xi_{QQ'}^*$ baryons are affected by the medium, considerably. Such that at   $\rho_N=1.5 ~\rho^{sat}_N$ their mass reach to $83\%$, $79\%$ and $77\%$ of their vacuum mass values for $\Xi^{*}_{cc}$,  $\Xi^{*}_{bc}$ and $\Xi^{*}_{bb}$ baryons, respectively. The negative shifts in the masses of these baryons represent their strong scalar attraction by the medium. We report the amount of the mass shift at each channel in table III. The analyses show that the masses of the $\Omega_{QQ'}^*$ doubly heavy baryons  are not affected by the medium. The vector self energy of the $\Omega_{QQ'}^*$ baryons are obtained to be roughly zero. However, the vector self energies of the $\Xi_{QQ'}^*$ baryons show considerable changes in the medium. The sign of the vector energies for $\Xi_{QQ'}^*$ reported in table III at saturated density are positive referring to considerable vector repulsion of these baryons by the medium. The residues of the $\Omega_{QQ'}^*$ baryons remain roughly unchanged by increasing the density of the nuclear medium, as well. However, the $\Xi_{QQ'}^*$ baryons' residues show drastic decrease with increasing the density of the medium. Note that, the dependencies of the parameters of the $\Xi_{QQ'}^*$ baryons on $\rho_N$ are  roughly linear. Their residues at $\rho_N=1.5 \rho^{sat}_N$  approach  to roughly $20\%$ of their vacuum values. The behavior of the parameters considered in the present study can be checked in future in-medium experiments. We turn off the density and obtain the mass of the baryons under consideration in $\rho_N \rightarrow 0$ limit (vacuum) and compare the obtained results with the previous predictions using different methods and approaches. We show our vacuum mass results and their comparison with other predictions in table II. We see over all good consistency of our results at $\rho_N \rightarrow 0$ limit with other approaches including the vacuum QCD sum rules within the errors. We hope that these results will help experimental groups especially at LHCb in the course of search for these baryons. We make our analyses in the interval $[0-1.5]~\rho^{sat}_N$ with $\rho^{sat}_N=(0.11)^3~GeV^3$ being corresponding to roughly $20\%$ of the density of the neutron stars' core. Our sum rules give reliable results within this interval.  One may extend the formalism to include higher densities to look at  the fate of these baryons at higher densities. This may help us in finding a critical density at which the doubly heavy baryons are melting. 

Production of the doubly heavy baryons  in cold nuclear medium needs simultaneous production of two pairs of heavy quark-antiquark  in medium. Then, a heavy quark from one pair requires  coming together with the heavy quark of the other pair, forming a heavy diquark of the total spin 1 or 0, previously discussed. The resultant heavy diquark requires  meeting  with a light quark to subsequently form  a doubly heavy baryon containing two heavy and one light quarks. These processes need that quarks be in the vicinity of  each other both in ordinary and rapidity spaces. We hope that future in-medium experiments will be able to provide these conditions required for the production of the doubly heavy baryons in nuclear medium. Our results may help the experimental groups in analyses of the future experimental data. Comparison of any data on the parameters considered in the present study with our predictions will provide useful information on the nature and internal structures of the doubly heavy baryons as well as their behavior in a dense medium. 
 
 \section*{Appendix : Spectral densities used in calculations}
 The spectral densities corresponding to the structures $g_{\mu\nu}$, $g_{\mu\nu}\pslash$ and $g_{\mu\nu}\uslash$ for the doubly heavy baryons $\Xi^*_{QQ}$ are obtained as
  \begin{eqnarray}
&&  \rho_{g_{\mu\nu}}^{QCD} (s, p_0)= \frac{1}{24576 \pi^{12}} \int^1_0 dz \int_0^{1-z}\frac{dw}{\big[w^2+(w+z) (z-1)\big]^5} m_Q \Bigg\{ \pi^2 \langle \frac{\alpha_s}{\pi}G^2\rangle_{\rho_N} \Big(w^2 +(w+z) \nonumber \\
&\times&(z-1)\Big) \Bigg[(w-1)w^3(3w-1)+7(w-1)^2w^2z+2(3w-2)z^4+\Big(13(w-1)w+1\Big)z^3+w\Big(w(13w \nonumber \\
&-&21)+7\Big)z^2+3z^5\Bigg] + 3(5w+3z)\Bigg[m_Q^2(w+z)\Big[z^2+(w-1)(z+w)\Big]-3swz(w+z-1)\Bigg]\nonumber \\
&\times&\Bigg[m_Q^2(w+z)\Big[z^2+(w-1)(z+w)\Big]-swz(w+z-1)\Bigg]\Bigg\} \Theta[L(s,z,w)] \nonumber \\
 &+&\frac{1}{3072 \pi^{6}} \int^1_0 dz \Bigg\{  \langle \bar{u}g_s\sigma G u\rangle_{\rho_N} (3+32z-32z^2)+4\Bigg[m_q\Big(4m_Q  \langle \bar{u}u\rangle_{\rho_N} -48 p_0 \langle u^{\dag} u\rangle_{\rho_N}(z-1)z\nonumber \\
 &-&3 \langle u^{\dag} iD_0 u\rangle_{\rho_N} \Big)+8 \Big(3s  \langle \bar{u}u\rangle_{\rho_N} (z-1)z+m_Q\big(p_0\langle u^{\dag} u\rangle_{\rho_N} (z-2)+\langle u^{\dag} iD_0 u\rangle_{\rho_N}\big)\Big)\nonumber \\
 &+&  \langle \bar{u} iD_0 iD_0 u\rangle_{\rho_N}(-8z^2+8z-3)\Bigg]\Bigg\} \Theta[\tilde{L}(s,z)],
\end{eqnarray}

\begin{eqnarray}
&&\rho_{g_{\mu\nu}\pslash}^{QCD} (s, p_0)=\frac{1}{294912 \pi^{12}} \int^1_0 dz \int_0^{1-z}\frac{w z dw}{(w^2+(w+z) (z-1))^6}  \Bigg\{ -4\pi^2  \langle \frac{\alpha_s}{\pi}G^2\rangle_{\rho_N}  (w + z - 1) \nonumber \\
&\times& \big[3 w^2 + w (5 z - 3) + 3 (z - 1) z\big] \big[w^2 + (w+z) (z - 1) \big]^2 -144 \Bigg[m_Q^2 (w+z)\Big(z^2+(w+z)(w-1)\Big)\nonumber \\
&-&swz(w+z-1) \Bigg] \Bigg[m_Q^2 \Big(z^2+(w+z)(w-1)\Big)\Big(z^2+w(w-1)+z(4w-1)\Big)-6swz(w+z-1)^2\Bigg]\Bigg\} \nonumber \\
&\times& \Theta[L(s,z,w)] + \frac{1}{576 \pi^{6}} \int^1_0 dz \Bigg\{-9 m_Q  \langle \bar{u}u\rangle_{\rho_N} +2\Big[5m_q\langle \bar{u}u\rangle_{\rho_N} -2 \langle u^{\dag} iD_0 u\rangle_{\rho_N} -3p_0 \langle u^{\dag} u\rangle_{\rho_N} \Big] \nonumber \\
&\times&(z-1)z \Bigg\} \Theta[\tilde{L}(s,z)],
\end{eqnarray}
and
\begin{eqnarray}
&&\rho_{g_{\mu\nu}\uslash}^{QCD} (s, p_0)=\frac{1}{1152 \pi^{6}}\int_0^{1-z} dz \Bigg\{  6 \big[3m_q-m_Q\big]m_Q \langle u^{\dag} u\rangle_{\rho_N}  + 2 \Big[8m_q p_0  \langle \bar{u}u\rangle_{\rho_N} -6 \langle u^{\dag} iD_0 iD_0  u\rangle_{\rho_N} \nonumber \\
&-& 32 p_0 \langle u^{\dag} iD_0 u\rangle_{\rho_N} +9 s \langle u^{\dag} u\rangle_{\rho_N}\Big] (z-1)z\Bigg\}\Theta[\tilde{L}(s,z)],
\end{eqnarray}
where the explicit forms of the functions $L(s,z,w)]$ and $\tilde{L}(s,z)$ are
\begin{eqnarray}
\label{ }
&&L(s,z,w)]=\nonumber \\
&-&\frac{(w-1) (m_Q^2 w (w^2+w (z-1)+(z-1) z)+z (m_ {Q}^2 (w^2+w (z-1)+(z-1) z)-s w (w+z-1)))}{(w^2+w (z-1)+(z-1) z)^2},\nonumber \\
&&\tilde{L}(s,z)= m_Q^2 (z-1)-z (m_ {Q}^2+s (z-1)).
\end{eqnarray}

\bibliography{refs}
\end{document}